\documentclass{emulateapj}

\journalinfo{Accepted for publication in  ApJ, December 1, 2008}

\shorttitle{Age/Mass Distribution in USco}
\shortauthors{Slesnick et al.}

\begin{document}

\title{A Large-Area Search for Low-Mass Objects in Upper Scorpius II: Age and Mass Distributions}

\author{Catherine L. Slesnick\footnote{current address: Dept. of Terrestrial Magnetism, Carnegie Inst. of Washington, Washington, DC 20015}, Lynne A. Hillenbrand \& John M. Carpenter}

\affil{Dept.\ of Astronomy, MC105-24, California Institute of
Technology,Pasadena, CA 91125}

\email{cslesnick@ciw.edu, lah@astro.caltech.edu, jmc@astro.caltech.edu}

\begin{abstract}

We present continued results from a wide-field, $\sim$150 deg$^2$, 
optical photometric and spectroscopic survey of the northern part of the $\sim$5
Myr-old Upper Scorpius OB Association.
Photometry and spectral 
types were used to derive effective temperatures
and luminosities and place newly identified association members onto a theoretical Hertzsprung-Russell 
diagram.  
From our survey, we have discovered 145 new low mass members of the association, 
and determined $\sim$10\% of these objects to be actively accreting material from a surrounding circumstellar disk.
Based on comparison of the spatial distributions of low and high mass association members, we find
no evidence for spatial segregation by mass within the northern portion of the association.  
Measured data are combined with 
pre-main sequence evolutionary models to derive a mass and age for each star.  Using Monte Carlo simulations
we show that, taking into account known observational uncertainties, the observed age dispersion for the low
mass population in USco is consistent with all stars forming in a single burst $\sim$5 Myr ago, and place an upper limit of $\pm$3 Myr on the age spread if
the star formation rate has been constant in time.
We derive the first spectroscopic mass function for USco that extends into the substellar regime,
and compare these results to those for three other young clusters and associations.
\end{abstract}


\section{Introduction}
\label{cha7:intro}

The review of \citet{1987ARA&A..25...23S} laid the basic framework for how isolated stars are created.
However, 
most stars do not form in isolation but rather in groups (i.e., clusters or associations; \citealt{1993AJ....105.1927G}, \citealt{1995AJ....109.1682L}, \citealt{2000AJ....120.3139C}), and many details of the processes involved
in stellar group formation
remain unexplained.  In particular, 
do all members of an individual 
group form in a single burst, 
or is star formation a lengthy process?  
The distribution of stellar 
masses formed within a stellar group is known  
as the initial mass function (IMF).  Is the IMF universal, or does it vary
with either star formation environment or time?

The phenomenon of rapid (1--2 Myr) clustered
star-formation has been found in almost all nearby young associations
(e.g., \citealt{2003ARA&A..41...57L}). 
The large numbers of very young stars and apparent lack of more evolved (5--10 Myr-old)
objects in star forming regions
contrasts with 
ages of a few tens of megayears (e.g., Blitz \& Shu 1980) inferred for molecular
clouds.
Either star-formation takes place for only a small fraction of the
cloud lifetime, molecular clouds themselves live only a few megayears (e.g., Hartmann et al. 2001),
or older group members have been missed in observational surveys which are often biased towards
stars that are still actively accreting or which possess an optically thick disk.
Observational surveys capable of detecting the entire age range present within a stellar group 
are needed if we are to begin to understand the time dependence of star formation
within a molecular cloud.

The basic structure of the field star IMF in the mass range
$\sim$0.5 M$_\odot$ to 10 M$_\odot$ 
is well characterized by a power law with 
slope $dN/dM \propto M^{-2.35}$, as derived originally by \cite{1955ApJ...121..161S}.
However, while the shape of the stellar mass function between
$\sim$0.5 M$_\odot$ and 10 M$_\odot$ is near universal,
below $\sim$0.5 M$_\odot$ the shape of the mass function 
may be strongly dependent on environment within the parental molecular
cloud (\citealt{2003ApJ...593.1093L}, \citealt{2004ApJ...610.1045S}).
Young stellar clusters and associations are particularly valuable for examining the 
shape of the low mass IMF because contracting low-mass pre-main sequence stars and brown dwarfs 
are 2-3.5 orders of magnitude more luminous than their counterparts on the main sequence. 
Thus, despite the fact that they are farther away, young stars and brown dwarfs can be more readily detected than stars of equivalent mass in the field.

The Upper Scorpius OB Association (USco) is the closest (145 pc; \citealt{1999AJ....117..354D}) young OB association to the Sun with  
120 members more massive than $\sim$1 M$_\odot$, and an associated population of $\sim$400 known
low mass objects (e.g., \citealt{1998A&A...333..619P}).  
Because of its poximity, and large population that spans the range from $<$0.02 M$_\odot$ to $>$10 M$_\odot$, 
USco is an ideal region in which to study the 
mass distribution of stars in OB associations. 
Furthermore, at an age of $\sim$5 Myr \citep{2002AJ....124..404P}, USco is young enough that we may be
able to measure the ages of member stars to a precision of less than a million years, and old enough that an on-going
episode of star formation lasting several millions of years could be detectable. 
A major difficulty faced by studies of the USco region is that
 the high mass members alone span $>$300 deg$^2$ on the sky. Obtaining a 
complete census of the association's low mass population is thus a formidable  
task as one must identify faint objects over a very large spatial region.  While there exist several techniques to identify young stars, 
many observational signatures (e.g., H$\alpha$ emission or near-infrared excess) probe accretion processes and circumstellar 
material rather than characteristics intrinsic to the stars themselves.

Recently, deep, multicolor imaging surveys combined with spectroscopic follow-up 
have proven successful in identifying 
low mass stars and
 brown dwarfs without active accretion,
as well as classical
T-Tauri type objects.
Young pre-main sequence (PMS) objects still undergoing contraction 
are cooler (i.e., redder in color) and/or more 
luminous (i.e., brighter in magnitude)
than their main sequence counterparts.  In nearby regions, candidate PMS stars can be identified through photometric colors and magnitudes
that are systematically
different from those of the bulk field star population. 
Spectroscopic follow-up observations allow assessment of surface gravity diagnostics which can be used to distinguish bonafide young PMS stars from foreground field dwarfs
and reddened background giants.
Previous imaging and spectroscopic surveys in USco include studies by \cite{2001AJ....121.1040P} and \cite{2002AJ....124..404P}, whose work
yielded 166 new low mass PMS objects.
\cite{2008MNRAS.383.1385L}, \cite{2004AJ....127..449M}, and \cite{2000AJ....120..479A}
together identified $\sim$70 members of USco with spectral type M6 or later.  
Thus far,
over 400 low mass ($M<$0.6 M$_\odot$) members have been identified in USco through X-rays, H$\alpha$ emission, photometry and/or spectroscopy. However, most searches have been limited to 
small subregions (smaller than a few square degrees) or bright objects ($R$ $\lesssim$ 16 mag). 
Given the derived mass function and number of observed high mass stars in USco, 
and assuming the high and low mass objects share the same spatial 
distribution, \cite{2002AJ....124..404P} estimate the entire USco region should 
contain $>$1500
young, low mass objects with $M<$0.6 M$_\odot$, most of which are yet to be discovered.

In Slesnick, Carpenter, \& Hillenbrand (2006a; hereafter Paper~I), we introduced our wide-field photometric survey covering $\sim$150 deg$^2$ of USco.  
In that work, we presented optical spectroscopic observations of 62 photometrically-selected candidate PMS stars, of which 43 were confirmed by us to be new USco members. 
In this companion paper we present spectra for an additional 178 candidates observed at either Palomar Observatory or Cerro-Tololo 
Inter-American Observatory (CTIO).
In \S\ref{cha:6:sec:obs}, we give a brief overview of the photometric and spectroscopic
surveys, and present new USco members discovered as a result of recent observations.  For the remainder of the paper beyond
\S\ref{cha:6:sec:obs},
we discuss results for new members presented here together with those for association members
identified in Paper~I.  
In \S\ref{cha:6:sec:discussion:sub:em}, we discuss H$\alpha$ emission profiles and accretion signatures in USco.
In \S\ref{cha:6:sec:discussion:sub:spat}, we compare the spatial distribution of the high and low mass populations.  
Finally, in \S\ref{discussion}, we construct an Hertzsprung-Russell (HR) diagram and present a discussion of the age and mass distributions derived for the 
members of USco presented in our work, 
as well as a comparison
of distributions derived for other nearby star forming regions.

\section{Observations}
\label{cha:6:sec:obs}

\subsection{Quest-2 Photometric Observations and Candidate Member Selection}
\label{obs:phot}

Driftscan observations were taken with the Quest-2 camera (\citealt{2003AAS...203.3812R}, \citealt{2007PASP..119.1278B}) on the 48-inch (1.2-m) Samuel Oschin Schmidt Telescope 
at Palomar Observatory.  Details of the observations and data analysis are given in Paper~I, Slesnick, et al. (2006b; hereafter SCH06), and \cite{mythesis}.
Three 4.$^\circ$6 wide scans, centered at $\delta$=-15.$^\circ$7, -19.$^\circ$5, and -23.$^\circ$3, were each observed between R.A. of 15h46m and 16h36m for a total survey area of
$\sim$150 deg$^2$.  The scan centered at $\delta=-19.5^\circ$ was observed
 3--4 times per night 
on seven consecutive, photometric nights between 2004 June 20-26.
The other two scans were observed once during this period.

\begin{figure}
\begin{center}
\scalebox{0.5}{\includegraphics{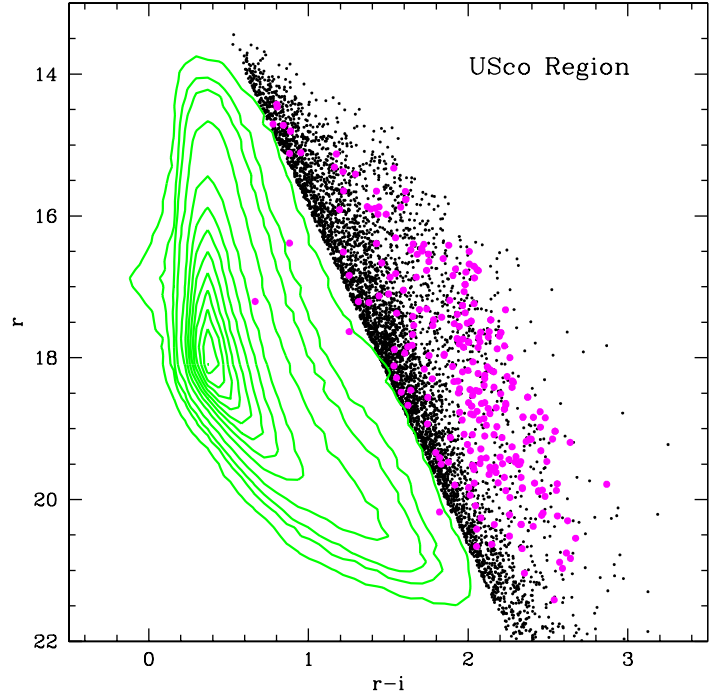}}
\caption[Selection of candidates from optical color-magnitude diagram]
{Optical color-magnitude diagram of all 
Quest-2 sources in the final USco catalog with $riJHK_S$ detections.
Contours represent the density of sources in the diagram, with
contour levels at 1\%, 2\%, 5\%, and 10--90\% of the peak value. 
Objects red-ward
of a linear approximation of the 1\% contour are shown as discrete points.
Objects for which we have spectral data 
are shown as large symbols. Photometry for four of the targets has changed significantly since
the first spectroscopic observations were taken (see SCH06) such that they would no longer be considered candidates.
}
\label{fig:cha3cmdselect}
\end{center}
\end{figure}

$B,R,I$ photometric data were calibrated to a system closely aligned with $Sloan$ $g,r,i$ magnitudes as described in SCH06 and \cite{mythesis}.
Magnitudes for all the new USco members identified in this study (based on a combination
of photometry and spectroscopy;
see \S\ref{cha:3:sec:summmmary}), both those newly identified in this work and those identified originally in Paper~I, are given in Table 1.  
Our goal is to use the photometry to isolate 
PMS stars 
from the field star population which dominates the $\sim$2 million member source catalog.
As discussed in \S\ref{cha7:intro}, nearby young stars 
occupy a sequence in an optical color-magnitude diagram (CMD) 
that is systematically brighter and/or redder than the sequence occupied by 
most field stars, and thus can be identified based on their colors and magnitudes. 
While the data have been calibrated to approximate $Sloan$ magnitudes [see \cite{mythesis} for more details], 
they are not precisely on any standard system, and we did not translate
theoretical isochrones into the Quest-2 system.
Thus, we 
did not select candidates based on isochronal data and instead,
considered as candidate PMS stars all sources redward of a linear approximation of the 1\% 
data contour in an $r,r-i$ CMD (Figure~\ref{fig:cha3cmdselect}).  
This criterion roughly corresponds to selecting all candidates
redward of a 30 Myr isochrone.

In addition to optical colors, the infrared 
colors and magnitudes of potential PMS candidates were also considered.
We matched the entire 
source catalog to the 2 Micron All Sky Survey (2MASS; \citealt{2006AJ....131.1163S}) and excluded 
from further consideration sources which did not have a 2MASS 
counterpart.
This criteria served to remove artifacts detected in the 2 driftscans for which we did not have repeated 
observations.
However, as discussed in Paper~I and SCH06, requiring a 2MASS detection biases the list of potential PMS candidates against faint blue sources. 
The position of each star 
on a
near-infrared color-color diagram was examined and any star with $J-H,H-K_S$ colors 
consistent with those of background
giants [$(J-H) > 0.6(H-K_S)+0.6$ or $(J-H) > 1.69(H-K_S)+0.29)$; \cite{1988PASP..100.1134B}] was excluded. We 
additionally considered $r-K_S$ colors 
and adopted the selection criterion $r < 2.57(r-K_S-3)+12.8$ as outlined in Paper~I.  
After 
all selection criteria were applied, the final candidate list 
contained $\sim$1700 candidate low mass PMS stars identified from $\sim$2 million sources detected in the 
Quest-2 survey. Of these stars, 90 correspond to previously known low mass members of 
USco (\citealt{2000AJ....120..479A}, \citealt{2002AJ....124..404P}, \citealt{2004AJ....127..449M}). 
The candidates were spread out over the entire $\sim$150 deg$^2$ survey area, which
encompasses $\sim$40\%
of the total spatial area spanned by the high mass members.  
Due to interstellar extinction and distance effects, optical and near-infrared colors and 
magnitudes alone are not a unique indicator of youth.  Therefore, it is necessary
to obtain spectroscopic follow-up observations for each photometrically-selected candidate.
This data allows for 
spectral type determination
and can confirm the presence of spectral features indicative of youth.

\subsection{Palomar Spectroscopy}
\label{palspec}
 
Moderate-resolution spectra of 105 PMS star candidates were taken with the Double 
Spectrograph on the Palomar 200-inch
telescope during the nights of 2006 May 16--19 and 2006 June 1-2.
All data were taken with the red side of the spectrograph through either the 1.5'' or the 
2'' slit using a 
5500 \AA$\;$
dichroic and a 316 lines mm$^{-1}$ grating blazed at 7500 \AA.  This set-up 
produced wavelength coverage from
6300 to 8825 \AA$\;$at a resolution of $R\sim$1250.  Typical exposure times were 
300--900 sec, and up to 1800 sec for the 
faintest targets ($r\sim$20).
Spectrophotometric standard stars \citep{1988ApJ...328..315M} were observed throughout 
each night
for flux calibration.  All sources were processed, extracted and calibrated 
using standard IRAF
tasks.  

\begin{figure}
\scalebox{0.4}{\includegraphics{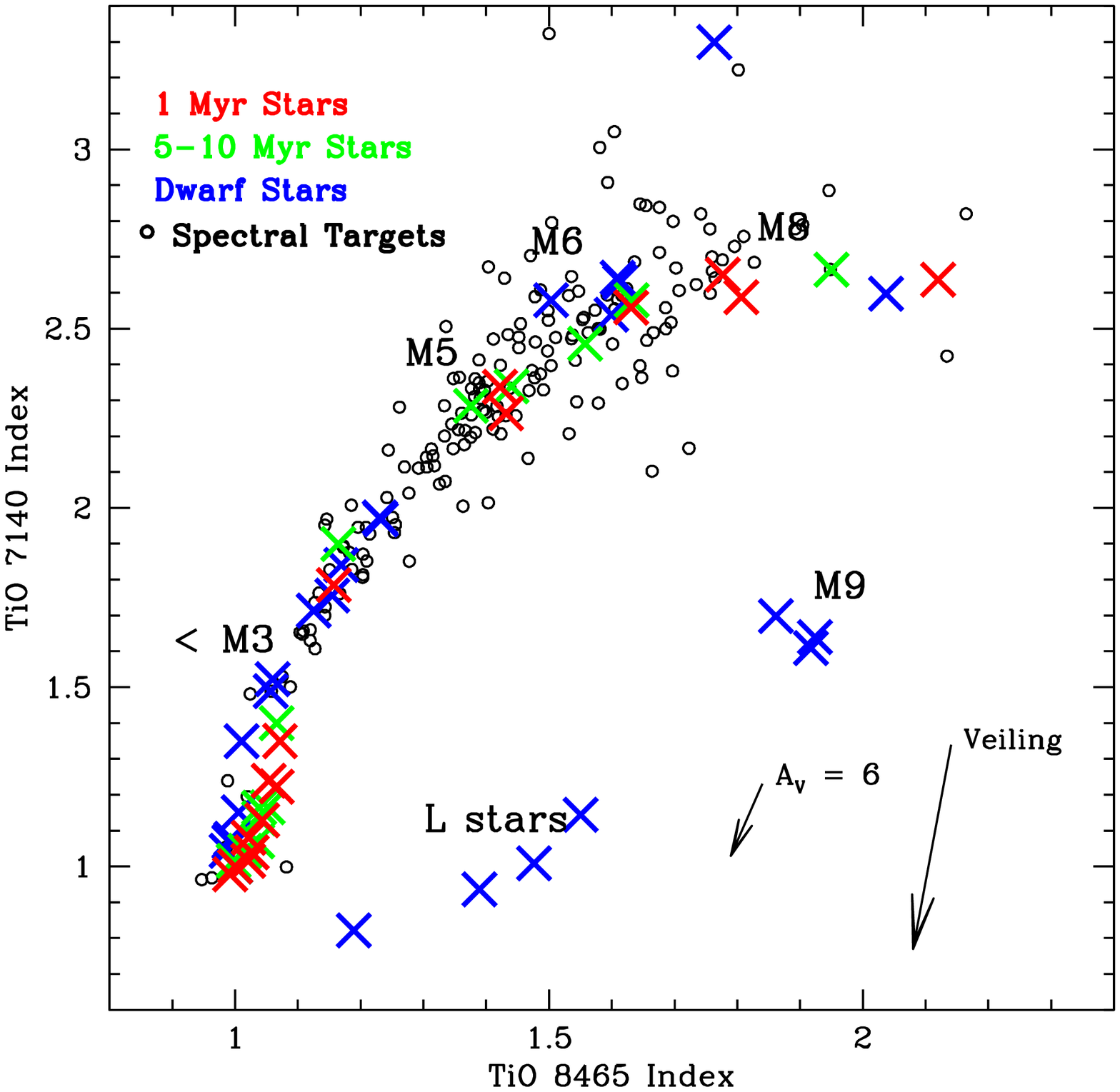}}
\hspace*{0.in}
\scalebox{0.4}{\includegraphics{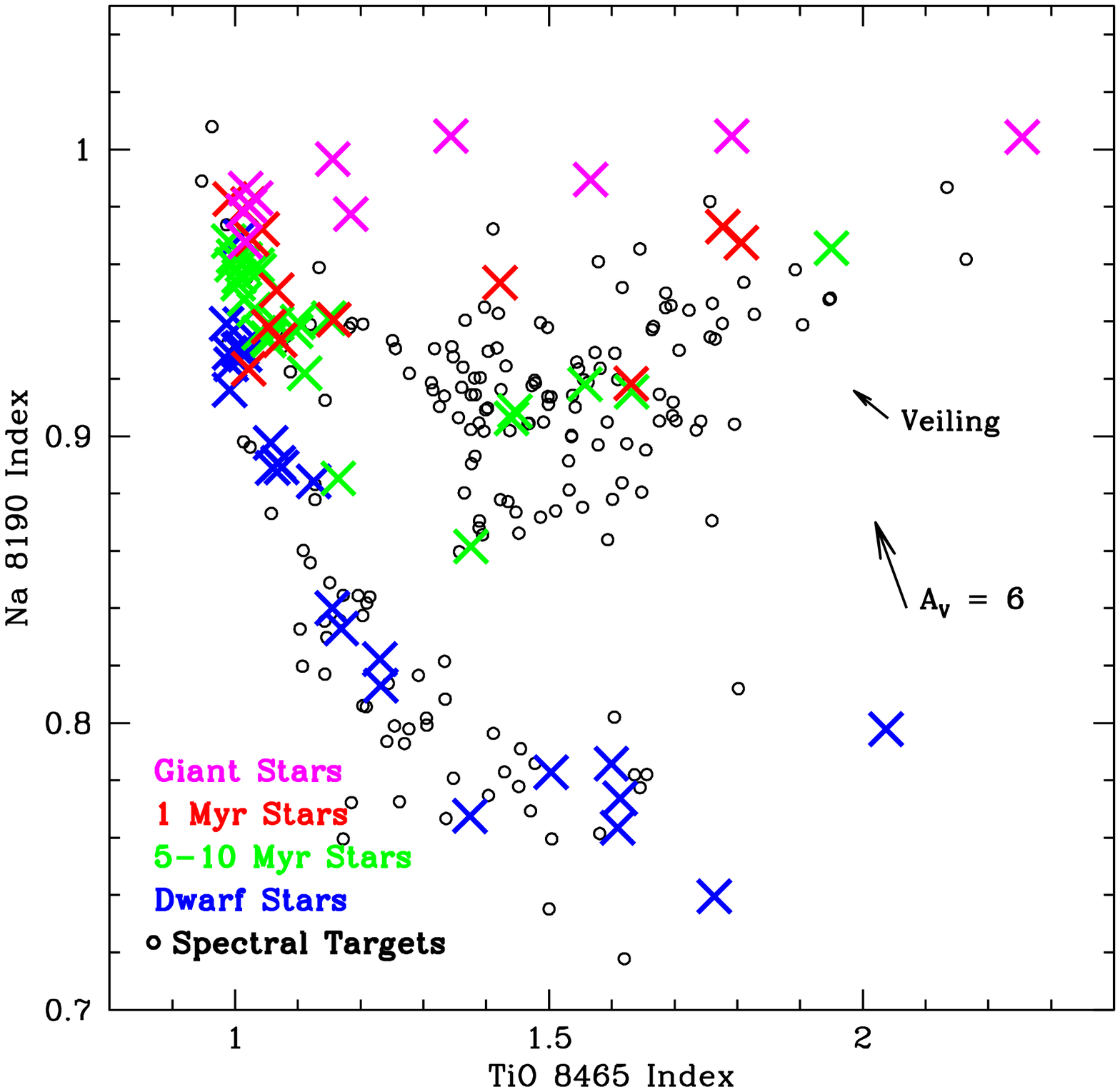}}
\begin{center}
\caption[Spectral indices for PMS candidates in USco]{Left panel shows 
temperature-sensitive TiO-7140 vs.~TiO-8465 indices; right panel shows 
TiO-8465 index vs.~gravity 
sensitive Na-8190 index.  In both panels, blue X's 
represent measured indices for a sample of old stars comprised of field dwarfs and
members of the Hyades ($\sim$650 Myr), Pleiades ($\sim$115 Myr) and AB Dor 
($\sim$75--150 Myr) associations.
Green X's show measured indices for intermediate-age spectral standards from 
Beta Pic ($\sim$11 Myr), TW Hya ($\sim$8 Myr),
and Upper Sco ($\sim$5 Myr).  Red X's show measured indices for young Taurus 
members ($\sim$1--2 Myr).  
Magenta X's in the right panel represent measured indices for giant standard stars.
In both panels, black circles are measured indices for USco PMS candidates observed at Palomar.  
The effects of 
extinction and veiling are shown as vectors (see Paper~I and SCH06). 
}
\label{fig:cha6uscoind}
\end{center}
\end{figure}

Spectral analysis for these observations were carried out as detailed in Paper~I, SCH06, and \cite{mythesis}.
Both spectral type (temperature) and surface gravity (age) determinations were first made 
through quantitative measurements of the TiO-7140, TiO-8465, and Na-8190 indices (defined in Paper~I) which measure
the strength of molecular and atomic absorption features present at optical wavelengths in the spectra of K- and M-type
stars. 
Figure~\ref{fig:cha6uscoind} shows spectral indices for 167 PMS spectral candidates in the USco region observed 
at Palomar (105 presented here for the first time and 62 presented in Paper~I). 
The left panel shows measurements of temperature-sensitive indices used to aid in spectral type determination.
We find fourteen outliers sit below the main locus of data 
points.
In all cases, the star is confirmed to exhibit low gravity signatures (see below) 
and we attribute the position in Figure~\ref{fig:cha6uscoind} to a small 
amount of veiling or reddening.
A detailed explanation of the possible effects of these processes
on the classification indices is given in Paper~I and SCH06. 

The surface gravity-sensitive Na-8190 index (Figure~\ref{fig:cha6uscoind}, right panel) allows us to easily separate the bona fide 
low gravity (i.e., young) objects 
from contaminant dwarfs stars 
over the spectral-type range $\sim$M3--M8.
We find a large fraction ($\sim$65\%) of the candidate 
objects have measured Na-8190
indices consistent with their having surface gravity less than that of field 
dwarfs 
at similar spectral types.  
The TiO and Na quantitative spectral indices were used to aid in classification only.  All final 
spectral type and gravity determination were done by visually comparing each spectrum to a grid
of spectral standard stars observed throughout the observing run.

\subsection{Cerro Tololo Spectroscopy}
\label{cha:3:sec:ctiospec} 

Additional spectra of PMS star candidates in USco were taken at 
CTIO using the Hydra multifiber spectrograph on the Blanco 4-m telescope during the nights of 2005 July 24--28.  In total, 
26 pointings were observed with the 
Site 2k$\times$4k CCD 
through the KPGLF grating.
The setup provided wavelength coverage from $\sim$6300 to 8660 \AA$\;$at a resolution
of 1.15 \AA/pixel. 

Fibers were placed preferentially on
stars meeting the criteria outlined in \S\ref{obs:phot}, thus providing 
a sample of candidates analogous to those observed at Palomar.  Remaining fibers 
were placed on stars meeting one or more of the criteria outlined in \S\ref{obs:phot}. 
For the practical requirements of the Hydra observations, stars were assigned to either `bright' pointings (corresponding to $r$ $\lesssim$ 16.5) or `faint' pointings
(corresponding to 16.5 $\lesssim$ $r$ $\lesssim$ 18.5).  Exposure times ranged from
900 to 1800 sec for bright pointings (dependent on weather conditions), and 
were set at 2700 sec for faint pointings.
In total, 10 bright and 16 faint pointings were observed during the 5 nights.
At each fiber 
configuration, we observed spectra of a comparison lamp and a quartz lamp 
to allow determination of the dispersion solution and throughput
correction during data reduction.  Biases and dome flats were taken each afternoon with 
all working fibers put in the
`large circle' configuration.
Milk flats (see below) were taken once during the observing run.

\subsubsection{Image Processing and Sky Subtraction}
\label{cha:3:sec:ctiospec:sub:proc} 

All frames were first pre-processed
(bias correction and trimming) in IRAF using the CCDPROC task.  Milk flats were obtained on the second afternoon
of observing.   This type of observation is an exposure of the daytime sky, and 
is taken through
a plate of milky glass placed between the output of the fibers
and the spectrograph camera.  The purpose of milk flats is to allow removal of CCD 
pixel-to-pixel variations from the data in the absence of fibers. 
After median-combining all observed milk flats, a spectral response image was created by smoothing in both X and Y directions.  The original combined milk
flat was divided by the spectral response image to 
create an image with a value of 1.0 everywhere except where pixel-to-pixel variations exist.
All data and calibration frames were divided by this image.  

Subsequent data reduction was done using the IRAF dohydra task.  All data for a given night were 
divided by the dome flat taken that afternoon.  Individual fiber-to-fiber throughput corrections (which change for each 
configuration) were made from division of 
each object frame by the corresponding quartz lamp spectra taken in the same
fiber configuration.  Wavelength calibration was carried out by matching each object spectrum to the corresponding lamp spectrum
observed through the same fiber.

For each pointing, a single sky spectrum was made using the sky generating tasks within dohydra.
Typically, $\sim$5--20 fibers were placed on sky during each observation.  The dohydra sky generating task first allows rejection of any
anomalous sky spectra which may have inadvertently fallen on a star or a clump of nebulosity.  
Remaining sky spectra are averaged
together using a 3$\sigma$ clipping algorithm.
The hydra field of view is 40' on a side.  
We found subtraction of a single sky spectrum for all spectra within a field often produced insufficient sky subtraction in that, for a given spectrum, all sky lines could be systematically
over- or under-subtracted.
To correct this problem, for each spectrum we computed the flux in a single sky line chosen to be towards the
center of the spectrum ($\sim$7316~\AA) and to be sufficiently isolated
that it was not blended with any other sky or stellar lines.
Object spectra were scaled 
such that the flux computed in the central sky line matched that of the sky
spectrum and the sky spectrum was subsequently  subtracted.

\subsubsection{Classification}
\label{cha:3:sec:ctiospec:sub:class} 

Spectral observations taken through fibers produce flux that varies as a function
of wavelength, dependent on the fiber configuration (i.e., how the relevant
fiber is bent and stretched to place it into position on the star).  Thus, flux correction of
fiber data is inherently difficult because it is not practical to observe a calibrator
star through every fiber at every configuration.  
We were therefore not able to flux-correct the hydra spectra in a manner analogous to the Palomar data
and could not
use the quantitative spectral indices (which rely on flux-corrected spectra) to aid in classification.
Though the overall spectral shape is not meaningful because the
spectra are not flux corrected, the depth of absorption and emission features is
not substantially affected.
During observations, 
we observed 
a range of known dwarf and giant stars (spectral types K0--M8).  In addition,
we observed known intermediate-age PMS members of the $\sim$30 Myr-old Tucanae
Hor association \citep{2004ApJ...612..496M} with early M spectral types.  
Spectra of previously identified K0--M8-type USco members were observed
in pointings towards USco along with program candidates.
All spectra of candidate PMS stars were classified by hand through 
comparison to each other and to standard stars observed with hydra. 

In total, the hydra observations yielded $\sim$1150 spectra (of varying quality). 
We first classified all of the spectra into broad categories.  
Approximately 450 were determined to 
be mid K to M `late-type' stars based on the presence of TiO molecular absorption in their spectra, 
and a lack of H$\alpha$ absorption. 
The largest constituent of the spectral sample ($\sim$650 stars) 
were `mid-type' stars with spectral types late F through
early K.  
These stars are characterized by a lack of TiO molecular 
absorption but the presence of Ca II triplet 
($\lambda$8498, $\lambda$8542 ,$\lambda$8662 \AA) and H$\alpha$ atomic absorption.
Absorption from the Ba II, Fe I, Ca I blend
at $\lambda$6497~\AA$\;$ 
begins to appear at late F/early G types 
and increases in strength compared to H$\alpha$ 
absorption through K spectral types.  Thirty of the late F -- early K type stars have H$\alpha$ emission present in their spectra.    
The remaining $\sim$50 stars observed with hydra were determined to have spectral types A--F with deep H$\alpha$ and Ca II triplet absorption
but noticeably lacking any absorption at the $\lambda$6497 \AA$\;$blend.  As expected, we saw no evidence of He absorption in
any of the spectra and conclude that our survey did not probe B- or O-type stars.

\begin{figure}
\begin{center}
\rotatebox{-90}{
\scalebox{0.5}{\includegraphics{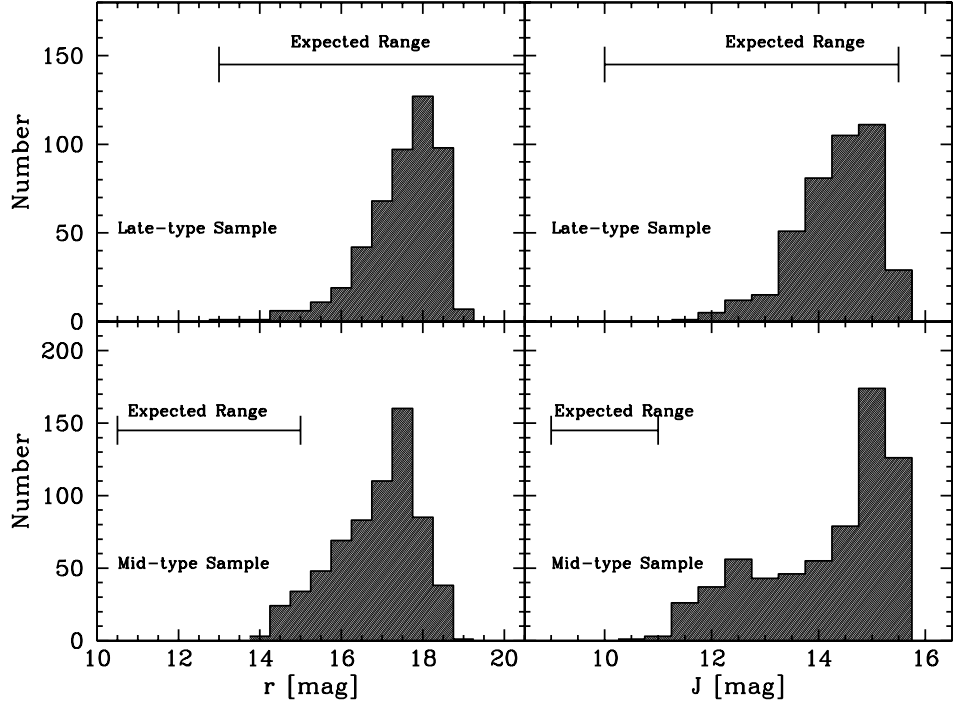}}
}
\caption[Distribution of magnitudes for Hydra targets]{Histograms of $r$- and $J$-band magnitudes for targets
of the Hydra spectral observations.  The top histograms show data for stars
classified as spectral type late-K to M; the bottom histograms show data classified as spectral type late F to early-K.
Shown in both panels is the expected range of magnitudes for 5 Myr-old stars of those spectral types at the distance of USco.
}
\label{fig:cha10hydrar}
\end{center}
\end{figure}

Magnitude histograms for the samples of observed `mid-type' and `late-type' stars are shown 
at $r$- and $J$-band (Figure~\ref{fig:cha10hydrar}),
along with the range of magnitudes expected for members of USco at these spectral types.   We have assumed for this
calculation that all members of USco are 5 Myr-old, 145 pc away, and have 0 $\leq$ A$_V$ $\leq$ 2 (Paper~I).  
As can been seen,
most of the stars classified as mid-type do
not appear to be members of USco based on their observed magnitudes and spectral types.  This sample is instead likely dominated
by reddened field dwarfs and background giants.  
This result is not surprising, considering most of these stars were observed through fibers placed on stars in the 
field of view that did not meet all 3 criteria for candidate selection (see \S\ref{cha:3:sec:ctiospec}  and \S\ref{obs:phot}),
and were not part of the $\sim$1700-member candidate list.

Thus far, we have derived detailed (at the $\sim$0.5 subclass level) classifications from these data
only for those stars whose colors and magnitudes meet all of the selection
criteria outlined in Paper~I and SCH06, thus providing a sample selected in exactly the same manner as the
stars observed at Palomar.  This sample was determined to consist entirely of late K- and M-type
stars, $\sim$50\% of which have low surface gravity consistent with association membership.
We present here results for these stars together with results 
from the Palomar spectral observations.

\subsection{Summary of Observations}
\label{cha:3:sec:summmmary} 

We identified a total of $\sim$2 million sources in the $\sim$150 deg$^2$ area covered by the Quest-2 imaging survey
of the northern part of USco.
The survey encompassed 56 out of 120 high mass association members identified with Hipparcos \citep{1999AJ....117..354D}, and $\sim$40\%
of the total spatial area spanned by the high mass members.
Using a combination of Quest-2 $g,r,i$ and 2MASS $J,H,K_S$ magnitudes, we selected $\sim$1700
candidate young stars based on placement in color-color and color-magnitude diagrams.
Specifically, to be considered as a candidate PMS star, the object was required to be within the reddest 1\% (in $r-i$ color) of stars observed in our survey, 
and to have infrared
colors inconsistent with field giants.

We obtained optical spectra for $\sim$15\% of the photometric candidates selected to be amoung the reddest of the photometric candidates
across all magnitudes (14$\lesssim r \lesssim$20) probed.
The goal of the spectroscopic observations is to measure spectral type
and confirm low gravity signatures consistant with bona fide PMS stars for each photometric candidate.
From the 105 Palomar spectral observations presented here (\S\ref{palspec}), we identified 66 new USco members with spectral
types ranging from M3 to M8.   We present an additional 36 new members with spectral types M3-M8 identified from CTIO observations. 
Quest-2 and 2MASS magnitudes for photometric candidates determined spectroscopically to be field dwarfs are given in Table~3. 

For the remainder of this paper, we will discuss together results from both the 102 new low mass USco members presented here,
and the 43 members presented in Paper~I.  Magnitudes and spectral measurements (i.e., spectral type,
H$\alpha$ equivalent width, and spectral indices, when applicable) for the 145 members identified by us are given in 
Table~1.  
Together, these stars represent a uniformly selected sample of new low mass USco members that is free
from bias in regards to circumstellar material and activity.

\section{Emission Line Objects}
\label{cha:6:sec:discussion:sub:em}

\begin{figure}
\begin{center}
\scalebox{0.45}{\includegraphics{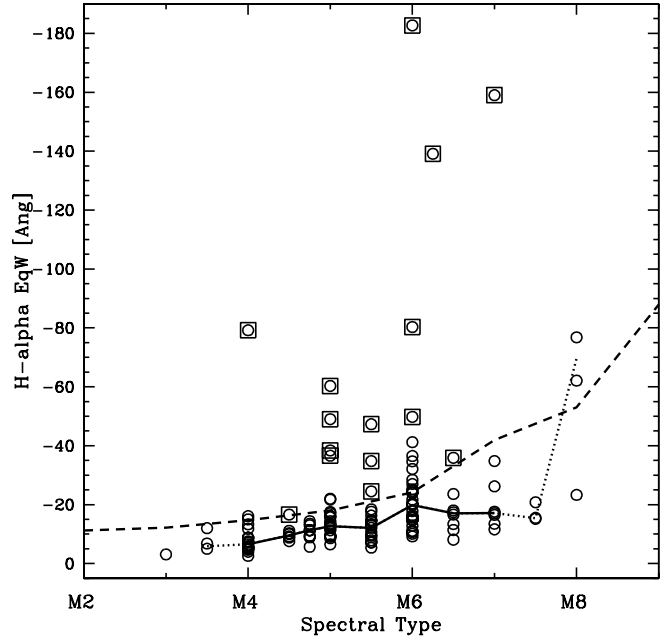}}
\caption[H$\alpha$ equivalent widths for PMS stars in USco]{Measured H$\alpha$ 
equivalent widths for all 145 low mass USco members presented here, shown as a function
of spectral type.  
Dashed line represents the empirical
accretor/nonaccretor upper limit derived by \cite{2003AJ....126.2997B}.  Solid 
line represents the median $W(H\alpha)$ for each spectral type.  
The spectral-type bins earlier than M4 and later than M7 
do not have enough measurements to derive a statistically representative
value for median H$\alpha$ emission
and we do not consider any
star outside the M4--M7 spectral type range to be in our sample of H$\alpha$ excess sources.  
Boxed objects have H$\alpha$ emission in excess of 3$\sigma$ above the median value for their 
spectral type, and
are considered by us to be actively accreting.
}
\label{fig:cha6halpusco}
\end{center}
\end{figure}

The most prominent emission line observed in the spectra of new members is H$\alpha$
which, seen in the spectra of young stars and brown dwarfs, is 
created via one of several mechanisms.
Weak, narrow H$\alpha$ lines are presumed to originate from active chromospheres 
whereas strong, broad and/or asymmetric lines can be produced from high-velocity, 
infalling accretion or strong winds.  
\cite{2003AJ....126.2997B} have proposed an empirical,
 spectral type-H$\alpha$ equivalent width ($W$(H$_\alpha$))
relation to describe the upper limit of non-accreting stars and brown 
dwarfs based on the chromospheric saturation limit observed in the Pleiades,
$\alpha$ Per, and IC 2391 open 
clusters.  These clusters are sufficiently old (50--125 Myr) that accretion should not be present,
and any observed H$\alpha$ emission is assumed to be produced entirely from
chromospheric activity.
Figure~\ref{fig:cha6halpusco} plots measured H$\alpha$ equivalent widths 
as a function of spectral type for the 145 members of USco presented here, 
shown with the \cite{2003AJ....126.2997B} empirical accretor/nonaccretor division.
Notably, every new member identified in our work shows H$\alpha$ in emission. 
Many stars and brown dwarfs
exhibit very strong H$\alpha$ emission  (see also Table~1)
at levels substantially above 
the accretor/nonaccretor division, 
and thus are possibly still undergoing active accretion. 

We determined an empirical criterion for identifying objects
with H$\alpha$ excess emission based on our data.
At an age of $\sim$5 Myr, we assume the bulk of our sample is no longer accreting and 
compute median values of H$\alpha$ emission as a function of spectral type
using 1-sigma clipping to remove any bias from outliers.  
For most bins, we define a star to have an H$\alpha$ excess if it exhibits emission at a level greater
than 3$\sigma$ above the median value for its spectral type, 
where $\sigma$ is the dispersion about the median for stars 
at a given spectral type.
Stars with such strong H$\alpha$ emission are likely to be accreting and are assumed such for the remainder of this work. 
These sources are distinguished on Figure~\ref{fig:cha6halpusco}.  For spectral type bins earlier than M4 and 
later than M7, we have identified fewer than 5 stars 
per spectral type.
Thus, for these bins we do not have enough measurements to derive a statistically representative
value for median H$\alpha$ emission and we do not consider any stars in these bins to be in our sample of H$\alpha$ excess, accreting sources. 
However, we note that two M8 stars (including the possible binary discussed in
Paper~I \S4.2) sit above the \cite{2003AJ....126.2997B} accretor/nonaccretor dividing line.

\begin{figure}
\begin{center}
\rotatebox{90}{
\scalebox{0.6}{\includegraphics{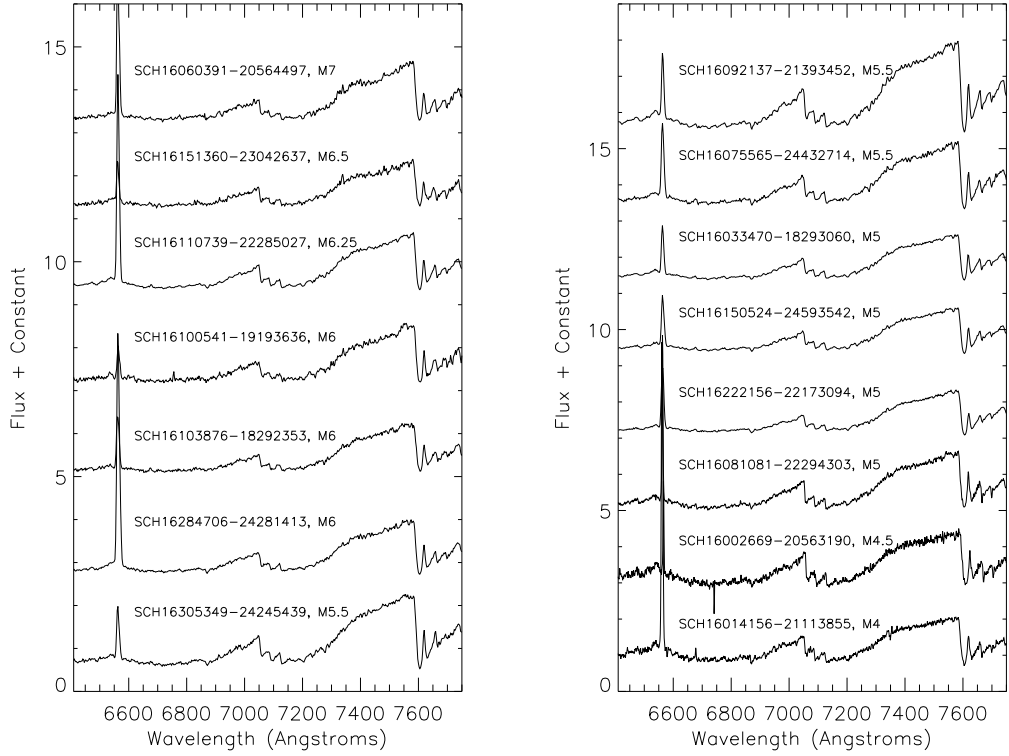}}}
\caption[Spectra of accreting stars]{
Spectra of the 15 stars determined to be accreting (as defined in \S\ref{cha:6:sec:discussion:sub:em}), 
shown in order of decreasing spectral type.} 
\label{fig:cha6hap}
\end{center}
\end{figure}

Based on the above criterion, we find 15 objects to exhibit H$\alpha$ emission with sufficient strength
to be considered by us to be actively accreting.  Spectra for accreting stars and brown dwarfs are shown
in Figure~\ref{fig:cha6hap}.
In addition to H$\alpha$ emission, many of these stars (SCH J16014156-21113855, SCH J16222156-22173094, SCH J16150524-24593542, SCH J16033470-18293060,
SCH J16075565-24432714, SCH J16284706-24281413, SCH J16103876-18292353, SCH J16110739-22285027, SCH J16060391-20564497) also exhibit
He I (6678 \AA) emission which is commonly seen in spectra of classical T-Tauri type objects.
Two accreting sources lie very close (within 
$\sim$1$^\circ$) to the young ($\lesssim$1 Myr) $\rho$ Ophiuchi molecular cloud ($\rho$Oph; $\alpha$=16 25 35.118, 
$\delta$= -23 26 49.84 J2000). However, because $\rho$ Oph 
lies slightly in front of USco \citep{2008ApJ...675L..29L}, if these stars were dynamically
ejected $\rho$ Oph members,
we would expect to see them exhibit systematically higher luminosities than USco members of similar spectral type.
Based on the derived HR diagram from these data (see \S\ref{cha:6:sec:discussion:sub:hr}) 
this phenomenon is not observed, and we include these two
stars in our sample of accreting members of USco.  We find at an age of $\sim$5 Myr (see \S\ref{cha:7:sec:usco}),
15/145 (or $\sim$10$^{+3}_{-3}$\%) of low mass association members (spectral type $\leq$M7 and $\geq$M4) 
are observed to be accreting based on 
the strength of H$\alpha$ emission present in their spectra.  If we were to use the \cite{2003AJ....126.2997B} accretion boundary
instead of our own empirical classification, we would have determined 23/145 (16\%) low mass stars and brown dwarfs
in our sample are still accreting. 
In comparison, \cite{2006A&A...446..485G} find $\sim$65\% (20/31) of 1 Myr-old low mass objects in the 
subclusters of Taurus to be actively accreting based on the strength of H$\alpha$ emission observed
in their spectra compared to the \cite{2003AJ....126.2997B} accretion boundary.  
Thus, a significant fraction of very low mass stars and brown dwarfs must
stop accreting between 1 and 5 Myr.   This conclusion is consistent with a median accretion lifetime
of $\sim$2--3 Myr for higher mass stars (\citealt{2001ApJ...553L.153H}, \citealt{2005astro.ph.11083H}).

\section{Spatial Distribution of Low Mass Stars}
\label{cha:6:sec:discussion:sub:spat}

\begin{figure}
\begin{center}
\scalebox{0.5}{\includegraphics{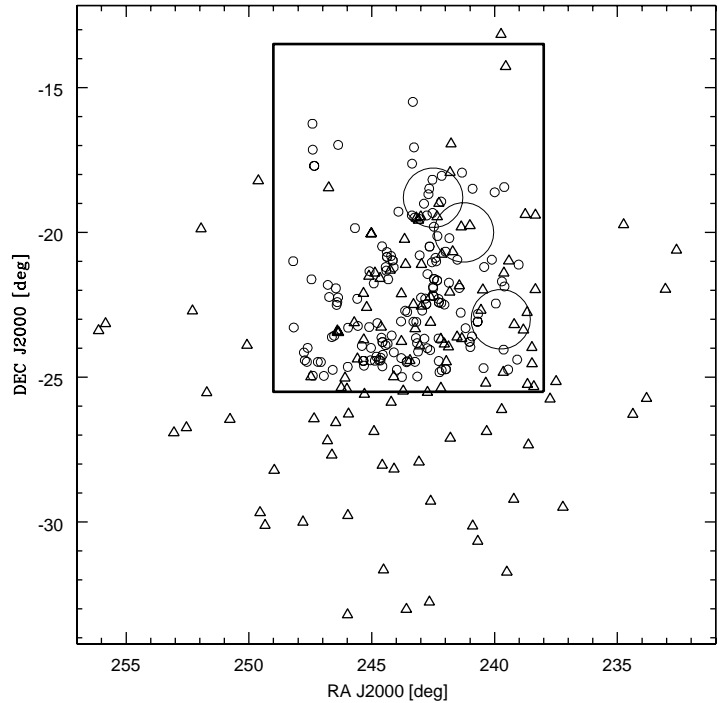}}
\caption[Spatial distribution of Hipparcos and Quest-2 stars in USco]{Spatial distribution of the 120 known Hipparcos members of USco
(open triangle) shown with stars observed spectroscopically by us that were determined to be new members (open circles). 
The Quest-2 survey area is boxed in black. The approximate location and size of the 2dF fields observed
by \cite{2002AJ....124..404P} are shown as large circles.}
\label{fig:cha6spath}
\end{center}
\end{figure}

Figure~\ref{fig:cha6spath} shows the 2D spatial distribution of the 120 high mass members
of USco identified 
in the Hipparcos survey \citep{1999AJ....117..354D}.  This sample represents the complete population of known members 
more massive than $\sim$1 M$_\odot$.  
The density of high mass stars is roughly constant from 237$\lesssim \alpha \lesssim$249$^\circ$ and peaks at $\delta \sim$-24$^\circ$.
As can be seen, despite the large area of the Quest-2 survey, it still encompassed only 
the central $\sim$13$^\circ$ in RA and the northern $\sim$12$^\circ$ in DEC of the association.

\begin{figure}
\begin{center}
\rotatebox{-90}{
\scalebox{0.45}{\includegraphics{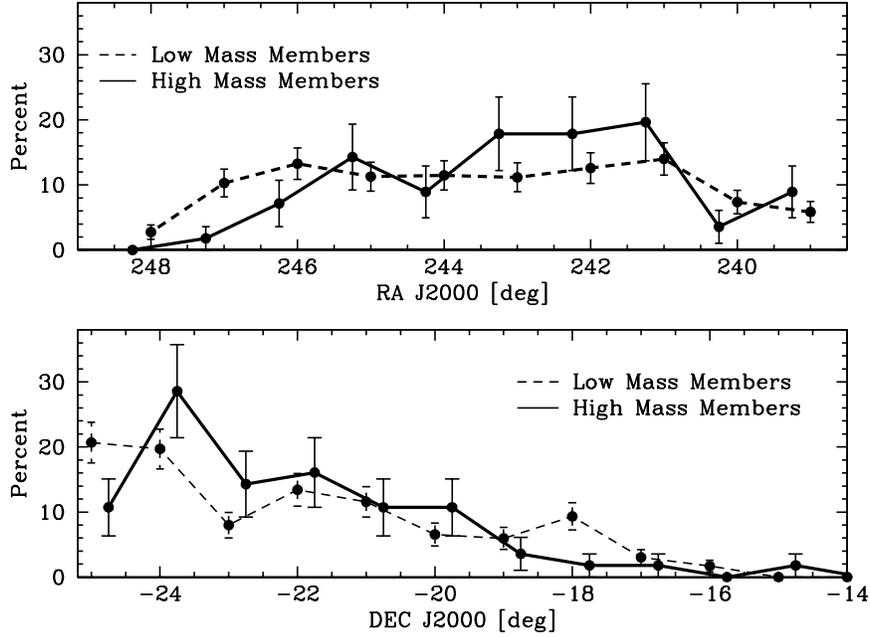}}}
\caption[1D spatial distribution of high and low mass stars in USco]{Top panel shows the percentage of
the low mass stars ($M\lesssim$0.6 M$_\odot$) discovered in this work (corrected to a uniform 25\% of candidates observed across all values) 
that lie at a given right ascension (dashed line) together with same information
for the 56 high mass Hipparcos stars ($M\gtrsim$1 M$_\odot$) found in the Quest-2 survey area (solid line).  Bottom panel illustrates the
same information as a function of declination.  In both
top and bottom panels, the coordinates of the high mass star bins have been shifted by 0.25$^\circ$ for ease of comparison between the 
high and low mass samples.  Errorbars assume Poisson statistics.} 
\label{fig:cha6spatp1}
\end{center}
\end{figure}

In general, the low mass PMS stars presented here share a common spatial distribution with the
high mass Hipparcos members.  Efforts to observe northwest of the Hipparcos
stars 
largely yielded reddened field dwarfs rather than young association members. 
To correct for bias in the spatial area we observed spectroscopically, we first computed the 
percent of photometric candidates that we observed spectroscopically, in 1-degree bins,
as a function of right ascension and declination.  
We found that we observed spectroscopically a maximum of $\sim$25\% of the photometric
candidates with a given degree-wide spatial bin.
We thus corrected 
every bin to a uniform 25\% of candidates observed, and calculated the number of members we would have detected assuming 
we had observed 25\% of the photometric candidates in each bin, 
and that the percentage of identified members relative to the number of 
stars observed spectroscopically at each spatial bin would remain unchanged.
Figure~\ref{fig:cha6spatp1} shows the resultant 1D spatial distributions for the low mass association
members discussed here, together with those for the 
56 Hipparcos stars that fall within 
our survey area.  
We conclude that the density of low mass association members found in the Quest-2 survey
is roughly uniform in RA, and peaks at $\delta \sim$-25$^\circ$ with stellar densities falling
off beyond these values.  
We find no evidence for spatial segregation by mass in USco within the northern portion of the association.

\section{Age and Mass Distributions}
\label{discussion}

\subsection{HR Diagram for New USco Members}
\label{cha:6:sec:discussion:sub:hr}

In this section, we
combine the spectral type and photometry of each new member to derive 
values for its luminosity
and effective temperature, and place it onto an HR diagram. 
As described in \S\ref{obs:phot}, the final Quest-2 photometry is not on a standard
magnitude system. 
Thus, because of the reliability and uniformity of the 2MASS
survey,
we chose to use $J$-band magnitudes and $(J-H)$ colors to derive luminosities.
An empirical fit to BC$_J$ as a function of spectral type was determined from the observational 
data of \cite{1996ApJS..104..117L} and \cite{2002ApJ...564..452L} (spectral types M1-M6.5 and M6-L3, respectively).
  We derived intrinsic colors, extinction, and 
effective temperatures using the methods described in 
\cite{2004ApJ...610.1045S}.  

\begin{figure}
\begin{center}
\scalebox{0.5}{\includegraphics{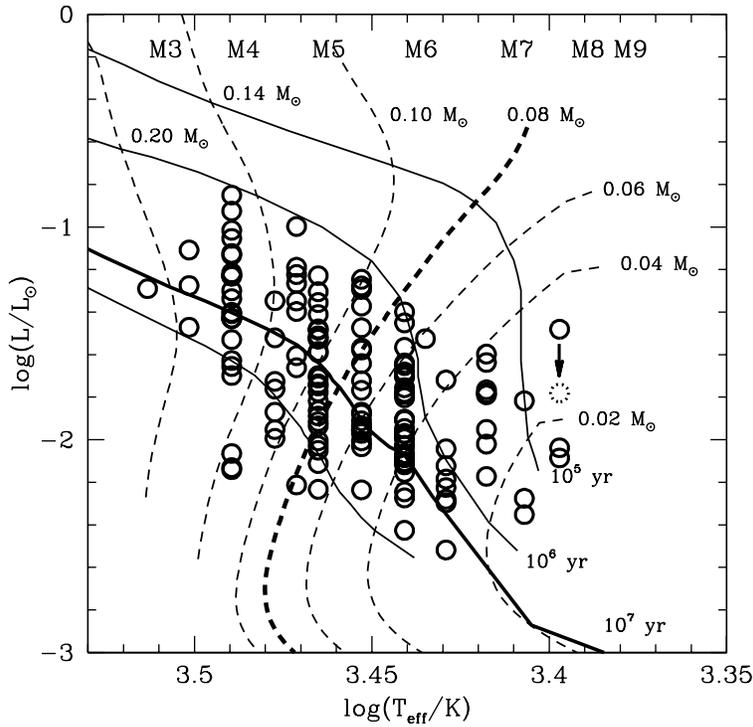}}
\caption[HR diagram for new PMS stars in USco] {HR diagram for new PMS objects found in the USco region, shown with model 
tracks and isochrones of \cite{1997MmSAI..68..807D}. The sample is consistent with an
age of $\sim$5 Myr and contains 
masses spanning the brown dwarf to stellar regimes.  The arrow plus dotted symbol indicates where the star SCH16224384-19510575 
would sit in the HR diagram as a single star if it is an unresolved, equal-mass binary (see \S4.2 in Paper~I).
}
\label{fig:cha6hrusco}
\end{center}
\end{figure}

In Figure~\ref{fig:cha6hrusco}, we present an HR diagram for the 145 low 
mass members of USco that we identified, shown with PMS model tracks and isochrones.
The most commonly used PMS models for low mass stars and brown dwarfs are 
those derived by \cite{1997MmSAI..68..807D} (hereafter DM97) and \cite{1998A&A...337..403B}, which differ primarily in their 
atmospheric approximations and treatment of convection.
Both models suggest similar mass ranges for our data of 0.02M$_\odot$ $< M <$ 0.2M$_\odot$,
though predicted masses for individual objects can vary by up to 0.09M$_\odot$ (60\%).
We have found that the slope of the DM97 isochrones provide a reasonable match to the derived HR diagram, 
whereas the 
\cite{1998A&A...337..403B} models predict systematically younger stars at lower masses.
A similar result was also noted by \cite{2008ASPC..384..200H}.  These authors compared the slope of six different theoretical PMS isochrones as a function of binary fraction, and found that, assuming an intermediate binary fraction consistent
with observations (e.g., \citealt{2008ApJ...679..762K}), the DM97 models provide the best match to the observed
HR diagram slope for stars in USco.  Thus, for the remainder of this work, we use the DM97 mass tracks and isochrones to derive mass and age for stars in our sample.  
All derived quantities are given in Table~2.

\subsection{Age Distribution of the Low Mass Population in USco}
\label{cha:7:sec:usco}

\begin{figure}
\begin{center}
\rotatebox{90}{
\scalebox{0.52}{\includegraphics{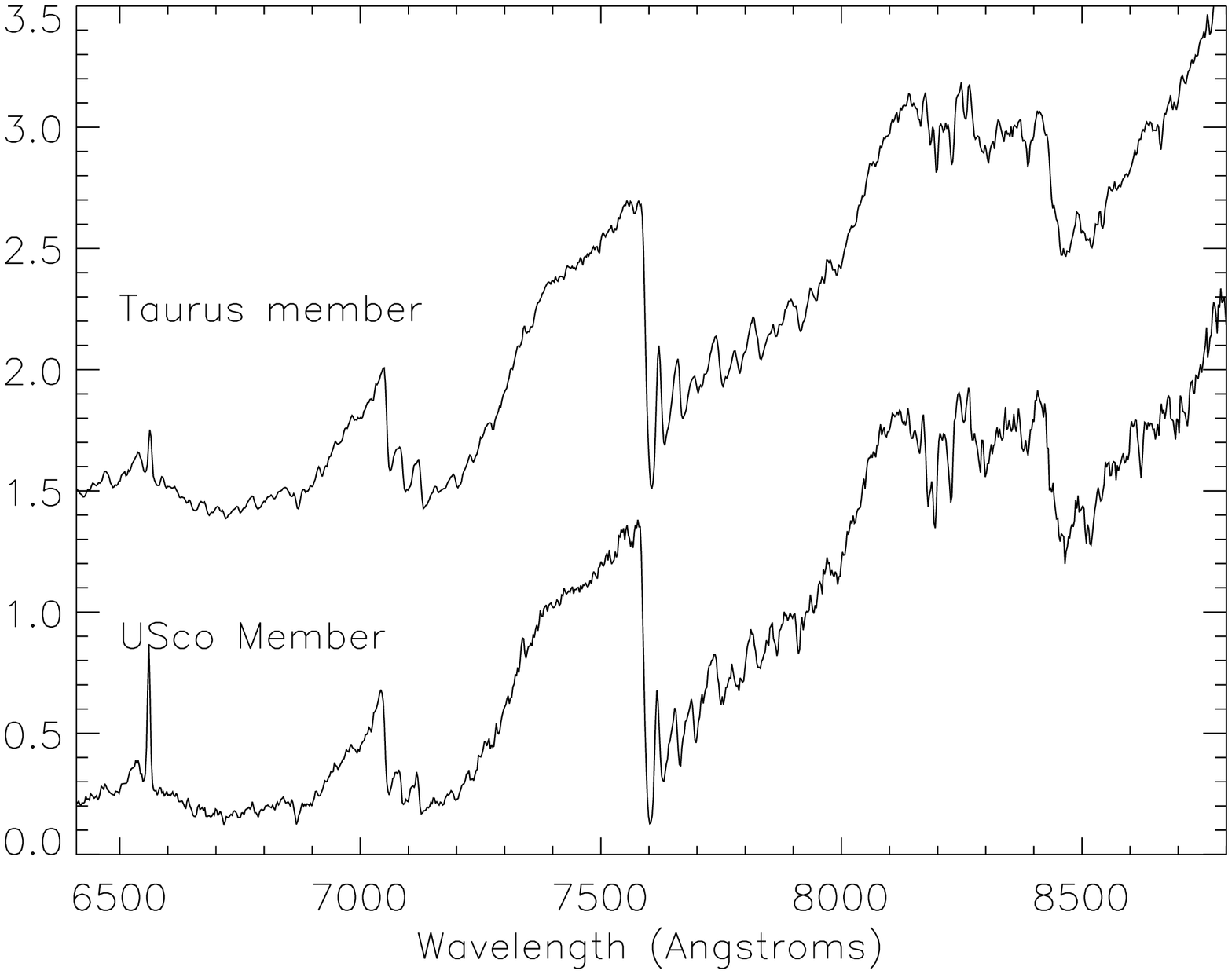}}
\hspace*{0.05in}
\scalebox{0.52}{\includegraphics{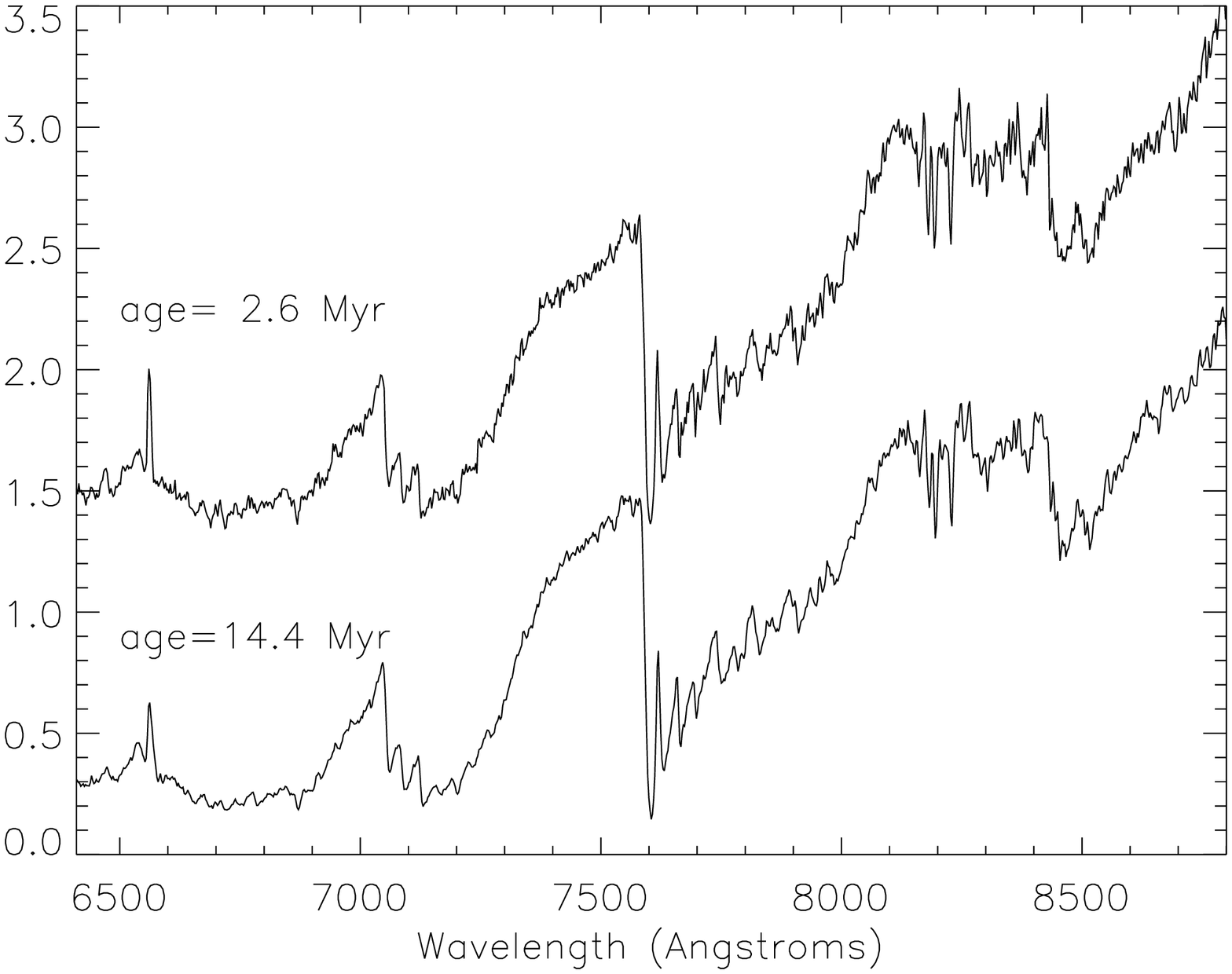}}
}
\caption[Spectra of an ``old'' and a ``young'' star in USco] {Top panel shows spectra of two stars with spectral type
M5 found in the USco survey.  The top spectrum is of the `youngest' (based on analysis of the HR diagram) 
M5 star observed spectroscopically 
at Palomar, $\sim$2.6 Myr-old;
the bottom spectrum is 
of the `oldest' (based on analysis of the HR diagram) 
M5 star observed at Palomar, $\sim$14.4 Myr.  However, these stars have near-identical spectra, and, based on analysis of
the strength of the surface gravity sensitive Na I line (8190 \AA), appear to be the same age.
In the lower panel we show spectra of known members of Taurus and USco, for comparison.  
The two stars have measured Na-8190 indices that are different by $\sim$9\%, and can be 
easily distinguished from each other visually.
}
\label{fig:cha7uscospec}
\end{center}
\end{figure}

Literal interpretation of the derived
HR diagram (Figure~\ref{fig:cha6hrusco})
reveals a population with median age of $\sim$4.1 Myr, and a continuous spread of ages over $\gtrsim$10 Myr.
While this result $may$ be real, the continuous
nature of the observed age distribution
in USco could also be produced from uncertainties in observed parameters.
Unlike the HR diagram, the observed surface gravity-sensitive spectral features do not show evidence for
a large spread in age between association members.  Figure~\ref{fig:cha7uscospec} shows spectra of two stars with spectral type
M5 identified in the survey.  The top spectrum is that of the `youngest' M5 star observed spectroscopically 
at Palomar (SCH16054416-21550566), $\sim$2.6 Myr-old
based on its location on the HR diagram (see also Table~2); the bottom spectrum is 
of the `oldest' M5 star observed at Palomar (SCH16162599-21122315), 
$\sim$14.4 Myr based on its location on the HR diagram.  
As discussed in SCH06 and \cite{mythesis},
young stars less than a few megayears old exhibit systematically less Na I ($\lambda$8190 \AA) absorption 
than do intermediate-age stars ($\sim$5--10 Myr) due to their lower surface gravity (see Figure~10 in Paper~I
and Figure~6 in SCH06).  Thus, if the derived ages from the HR diagram are
correct, the spectrum of the 14.4 Myr-old star should have noticeably stronger Na I absorption. 
For example, an M5 star at 1--5 Myr would have a Na-8190 index 5--10\% larger than that observed for an M5 star at 10 Myr.
The spectra presented in Figure~\ref{fig:cha7uscospec} are nearly identical (and have near-identical measured Na-8190 indices of 0.89 and 0.90).  Based on analysis of spectral features, we classified these two stars as being roughly the same
age.  
However, based on the differences in observed luminosity, interpreting the HR diagram literally,
one would infer an age spread of $>$10 Myr between the two stars.

\subsubsection{Comparison of the Observed Age Distribution to a Coeval Population} 
\label{cha:7:sec:usco:coeval}

Due to the discrepancy between stellar ages derived from the HR diagram compared
to those inferred from spectral surface gravity signatures, 
we sought to determine the statistical significance of the observed spread in ages for stars on the HR diagram. 
Effective temperatures and luminosities shown in Figure~\ref{fig:cha6hrusco} are derived from observed
$J$ magnitudes, $J-H$ colors, and spectral types.
The age and mass of a star are inferred from effective temperature and luminosity using a set
of theoretical isochrones and mass tracks.  Thus, uncertainties in 
measured photometry or spectral type 
are propagated into uncertainties in mass and age.
Variations in distance and binarity can cause additional uncertainty in luminosity, and, 
hence, quantities derived from the HR diagram.
 
We explore first the possibility that the USco population could be 
coeval, and the apparent age spread in the HR diagram is a result of observational uncertainties
combined with association depth and binarity effects. 
Using a similar method to that used in this work, \cite{1999AJ....117.2381P} compute an age of $\sim$5 Myr for the intermediate-mass
members of USco.  
We used Monte Carlo techniques to generate a coeval population of 5 Myr-old 
stars and brown dwarfs at a distance of 145 pc.
The input spectral type distribution was selected to mirror that of our observed distribution of stars, with 
1000 stars simulated for every star observed.  
For each star in the simulated population,
$J$- and $H$-band magnitudes were varied by adding random offsets drawn from a Gaussian 
distribution with a 1-sigma deviation of 0.025 mag, corresponding
to the average uncertainty in the 2MASS photometry for observed stars.  To mimic the magnitude-limited
data sample, we did not allow any star to be simulated below the photometric survey limits 
($J$=16 and $H$=15.5).  Similarly, a random offset was added to
the assumed spectral type, selected from a 
Gaussian distribution with 1-sigma errors of 0.5 spectral
subtypes corresponding to the qualitative error of the optical spectral type determinations.   
Simulated spectral types were rounded to the nearest 0.25 subclasses to reflect the discreet nature of 
spectral type classification for new members discovered in our work.
Using the new spectral type,
for each simulated star, 
we re-derived
the expected effective temperature, bolometric correction and intrinsic $J-H$ color
using the methods described in \cite{2004ApJ...610.1045S}.

The maximum distance spread (derived from secular parallax measurements) 
among members of the association with Hipparcos measurements 
is 50 pc (\citealt{2002AJ....124..404P}; \citealt{1999MNRAS.310..585D}).  In the simulation,
we assumed a uniform spatial distribution over a box of this depth centered at 145 pc.  
A 33\% binary fraction for stars across all simulated spectral types (M3--M8) was assumed, consistent with observational
results of the binary frequency for low mass members of USco (\citealt{2008ApJ...679..762K}, \citealt{2005ApJ...633..452K}).
All observed low mass binaries in USco have near equal masses ($m_{secondary}/m_{primary} \gtrsim$0.6; \citealt{2005ApJ...633..452K}), and
thus, 
the assumption was made that
all binaries were composed of two equal mass stars.  This assumption is somewhat liberal in that
it will produce the largest possible dispersion in luminosity.

\begin{figure}
\begin{center}
\rotatebox{-90}{
\scalebox{0.5}{\includegraphics{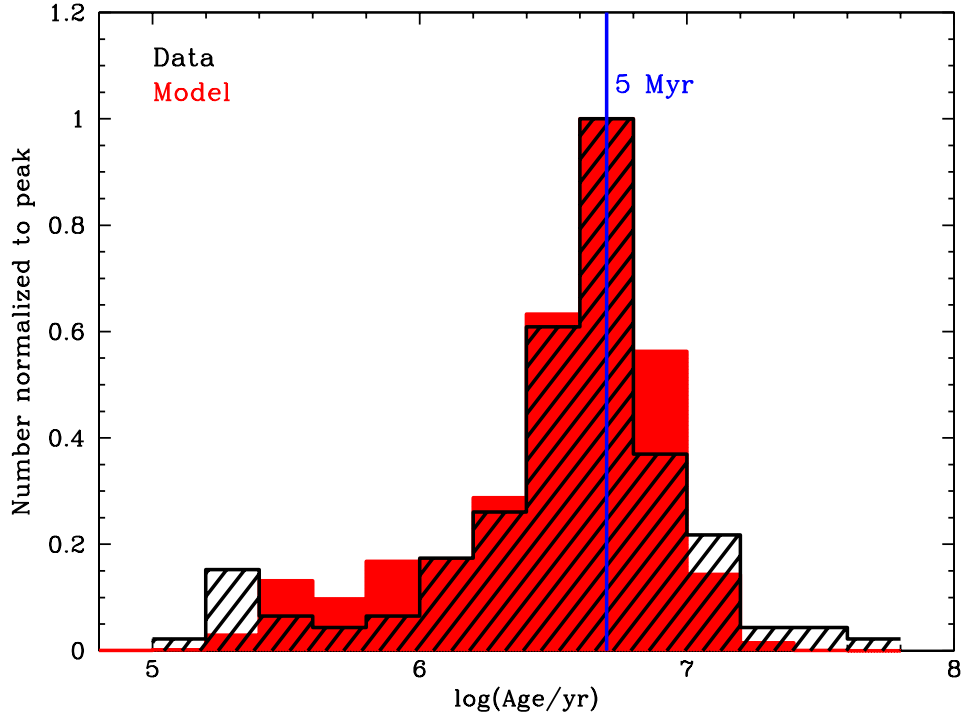}}}
\caption[Distribution of ages in USco inferred from the HR diagram] {Red shaded histogram shows the age distribution
derived from the 
Monte Carlo simulation, overplotted with a histogram of ages for the data (black hatched histogram).  
Both histograms have been normalized to 
unity at the peak for comparison, and stars having log($T_{eff})<$3.4 (beyond which interpolation of the isochrones
becomes unreliable) have been excluded from the data and the model results.  
The widths of the distributions are remarkably similar, given the simplistic nature of the Monte Carlo simulation. 
}
\label{fig:cha7ageusco}
\end{center}
\end{figure}

We compare quantitatively results of this simulation to the observed data in Figure~\ref{fig:cha7ageusco}.  The red shaded histogram shows the age distribution
derived from the 
Monte Carlo simulation, overplotted with a histogram of ages for the data (black hatched histogram).  
Both histograms have been normalized to 
unity at the peak for comparison, and all stars with log($T_{eff})<$3.4 (beyond which interpolation of the isochrones
becomes unreliable) have been excluded from the data and the model results.  
The widths of the distributions are remarkably similar, given the simplistic nature of the Monte Carlo simulation.  For the simulated association, we find a mean apparent age of log($age$)=6.51$\pm$0.40 whereas for the 
data (not including the 3 M8 stars) we find a mean age of log($age$)=6.53$\pm$0.47.
The data and the model differ significantly at the distribution tails, which is most likely caused either by an incorrect assumption of
one of the model parameters, or by an inherent problem with the theoretical isochrones at very low masses \citep{2004ApJ...604..741H}. 
To quantitatively assess how well the model reproduces the bulk of the data, excluding the distribution tails, we applied
a $\chi^2$ test of the central peaks of the distributions from log($age$)=6.0 to log($age$)=7.4.  This test yields 
a $\sim$53\% probability
that the two distributions could have been drawn from the same population.  
Thus, given the uncertainties, the data are consistent with most stars in USco
forming via a single burst $\sim$5 Myr ago.  

This result is somewhat surprising given
the large extent of the association ($\sim$35 pc across).
Assuming a sound speed of $c_S$=0.2 km/s consistent with a T=10 K molecular cloud, and assuming the stars formed close 
to where they are observed today, the sound crossing
time for the parental molecular cloud is $\sim$85 Myr.
Thus, one end of the cloud could not have `communicated' to the other in time to create a simultaneous burst
of star formation.  This result does not, however, rule out the possibility 
that star formation happened simultaneously throughout
the extent of the cloud because every part independently reached the threshold for star formation at the same time. 
Another scenario is that the stars formed much closer together and have since spread to their
current positions.  However, this possibility can be ruled out from simple arguments.  
The velocity dispersion of the massive Hipparcos members is $\sim$1.3 km/s
\citep{1999MNRAS.310..585D}.  In 5 Myr, the furthest an association member 
could travel at a speed of 1.3 km/s is
$\sim$6.5 pc, less than half the radius of the current association.

A similar discrepancy between a small observed age spread and the large
spatial extent of the association has also been noted in the literature for USco's intermediate and high mass members \citep{1999AJ....117.2381P}.  
To explain the disparity, \cite{1999AJ....117.2381P} (and later \citealt{2007IAUS..237..270P}) proposed a
scenario in which star formation in USco was triggered by an external event in the form of 
a supernova explosion
in the neighboring Upper Centaurus-Lupus association ($\sim$70 pc away and $\sim$17 Myr-old).  
They argue based on the structure
and kinematics of large H I loops surrounding Sco Cen, that such an event is evidenced to have occurred $\sim$12 Myr ago.
If true, the explosion would have driven a shock wave that would have reached USco about $\sim$5 Myr ago, consistent with the 
inferred 
age of USco's stellar population.   
Our results may imply a 
similar small age spread with an association age $\sim$5 Myr and large spatial extent (see \S\ref{cha:6:sec:discussion:sub:spat}) 
for USco's lowest mass stars and substellar members.
While our results do not prove 
the \cite{1999AJ....117.2381P} hypothesis true, they do 
give support to the hypothesis'
plausibility, and extend its validity to even the lowest mass association members.

\subsubsection{Comparison of the Observed Age Distribution to a Uniform Distribution} 
\label{cha:7:sec:usco:distrib}

From the simulation discussd in \S\ref{cha:7:sec:usco:coeval}, 
we have determined the observed age distribution in USco is consistent with all stars forming in a single burst
5 Myr ago.  We explore also the maximum age spread that can be inferred from
our data assuming that the star formation rate has been constant in time.
We repeated the 
original Monte Carlo simulation allowing age to vary in addition to spectral type, photometric error,
distance and binarity.  We began with a 5 Myr population, and for each star added a random
offset in age drawn from a uniform population between $\pm$0.5 Myr, $\pm$1 Myr, $\pm$1.5 Myr, $\pm$2 Myr,
$\pm$2.5 Myr, $\pm$3 Myr, $\pm$3.5 Myr, $\pm$4 Myr, $\pm$4.5 Myr, or $\pm$5 Myr. This age offset had the effect of changing the 
starting $J$ and $H$ magnitudes.
All other parameters were computed as described in \S\ref{cha:7:sec:usco:coeval}.

\begin{figure}
\begin{center}
\rotatebox{-90}{
\scalebox{0.5}{\includegraphics{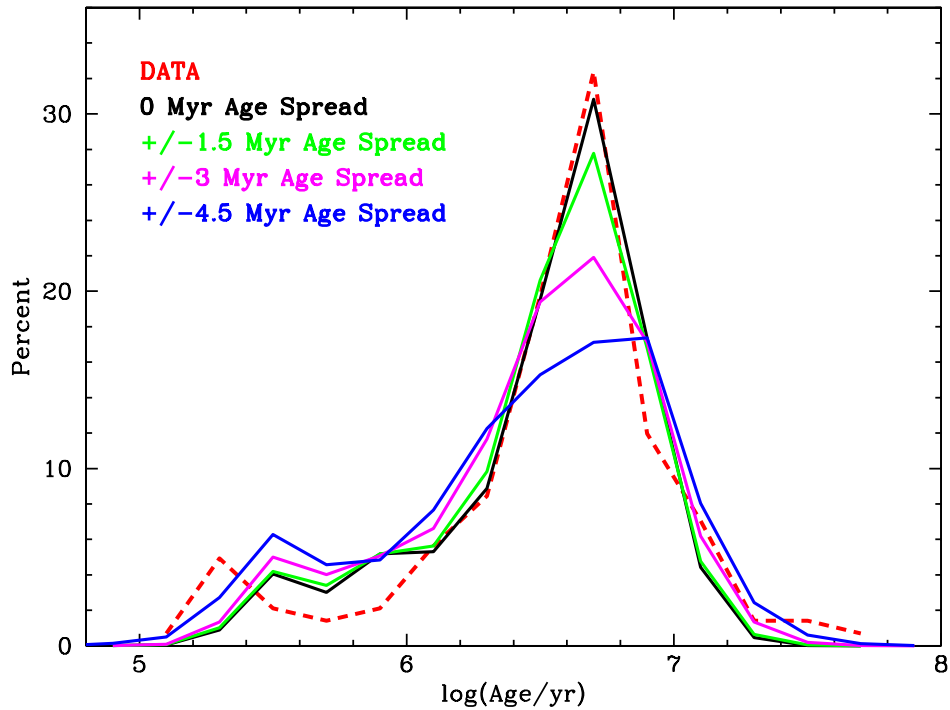}}}
\caption[Distribution of ages in USco inferred from model populations] {Red dashed line shows the age distribution
derived from spectral data, plotted as percentage of the total stars that are at a certain age.  Colored solid lines
show the same information for results form the coeval model (black), together with results from model age distributions of
5$\pm$1.5 Myr (green), 5$\pm$3 Myr (magenta), and 5$\pm$4.5 Myr (blue).  As expected, the peak of the simulated distribution decreases
and more power is seen in the wings as a larger age spread is injected into the population.
}
\label{fig:cha7ageuscomod}
\end{center}
\end{figure}

Figure~\ref{fig:cha7ageuscomod} compares the observed age
distribution derived from the HR diagram with the results from
the Monte Carlo simulation.  
As expected, the peak of the simulated distribution decreases
and more power is seen in the wings as a larger age spread is injected into the population.
We have run a $\chi^2$ test between central peaks of the data and the simulated model distributions between log($age$)=6
and log($age$)=7.4.  We find that the computed $\chi^2$ probability remains high, between 50\% and 60\%, for simulations
with an age spread of $\leq \pm$2 Myr, and then falls off rapidly with probabilities of $<$5\% beyond $\pm$3 Myr.  
Thus, we conclude that the observed low mass population of USco formed in $<$6 Myr, and most likely $<$4 Myr;
this finding is much less than the $>$10 Myr
age spread implied by literal interpretation of the HR diagram.

\subsection{The Low-Mass IMF}
\label{cha:7:sec:usco:sub:imf}

In this section we use model tracks to derive the first spectroscopic mass function for stars
and brown dwarfs in USco less massive than $\sim$0.1 M$_\odot$.
The study by \cite{2002AJ....124..404P} derived an IMF for the $stellar$ population in the association
by combining the 166 spectroscopically confirmed low mass members 
($M$$\sim$0.8--0.1 M$_\odot$) discovered via their 2dF survey with
the 120 known Hipparcos members (constituting a complete
sample of members more massive than $\sim$1 M$_\odot$)
and 84 X-ray selected members $(M\sim$0.8--2 M$_\odot$; \citealt{1999AJ....117.2381P}). 
The resultant mass function yielded a 3-segment power law function that was consistent with recent
field star and other cluster IMF determinations above $\sim$2 M$_\odot$, but with a 2$\sigma$ excess of stars above numbers seen in the field at lower masses.

The only previous survey that has attempted to derive an IMF for USco's substellar members 
is the photometric study by \cite{2007MNRAS.374..372L}.  These authors 
observed $\sim$6.5 deg$^2$ of USco at $ZYJHK$-bands as part of the UKIRT Infrared Deep Sky Survey (UKIDSS) Galactic Cluster Survey.  From these data, the authors used the $J$-band luminosity function 
to derive a $photometric$ IMF in USco from 0.3--0.007 M$_\odot$.  
They derived a mass function that was significantly flatter ($dN/dM \propto M^{-0.3\pm0.1}$) than the IMF derived by 
\cite{2002AJ....124..404P} for similar mass ranges ($dN/dM \propto M^{-0.9\pm0.2}$ for 0.1$\leq M$/M$_\odot < $0.6).
However, as noted in \cite{2004ApJ...610.1045S}, 
photometrically derived IMFs 
suffer from degeneracies between mass, age, distance, extinction, and photometric excess present in the CMDs
from which they are derived, and 
they cannot directly (only statistically) correct for field star contamination.

As was discussed in \S\ref{cha:7:sec:usco}, deriving
quantities from spectroscopic data placed onto an HR diagram has its own set of uncertainties.  We have already shown that known
uncertainties in observable magnitudes and spectral types can produce an apparent age spread
of $>$10 Myr, and therefore  
must also determine if uncertainties in the data can produce a false spread in the observed mass distribution.  
To address this issue, we derived
a theoretical IMF for a coeval population of stars at 5 Myr.
We assumed a spectral type distribution consistent with that
for our observed population, and assigned each star the mass that it would have at that
spectral type if it were 5 Myr-old.
Examination of the difference between the two mass distributions using a KS test
reveals that the mass distributions derived from the data and the theoretical
5 Myr-old population are statistically consistent with each other (probability=12\%).
Because the mass tracks are close
to vertical for low mass stars (see Figure~\ref{fig:cha6hrusco}), the dominant source
of uncertainty in derived masses will come from uncertainties in temperature, arising
from uncertainties in spectral type   
($\pm$0.5 subtypes).
Presuming we 
are equally likely to misclassify a star 0.5 subtypes too early as we are to misclassify it 0.5 substypes too late,
the mass distribution will not change significantly.  
Thus, we conclude that the mass distribution derived from the HR diagram is robust to observable uncertainties.
We note, however, that the mass distribution is strongly affected by systematics in the theoretical
models, and results will vary dependent upon which set of mass tracks are used.

\begin{figure}
\begin{center}
\rotatebox{-90}{
\scalebox{0.4}{\includegraphics{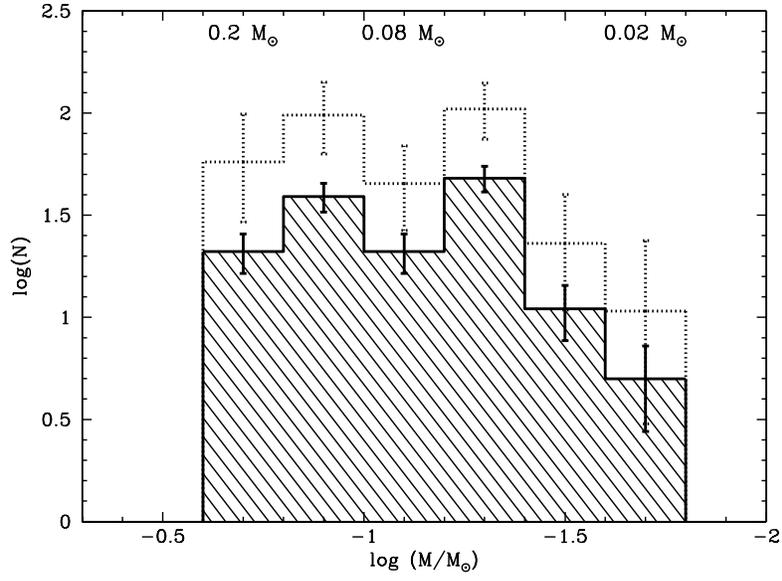}}}
\caption[USco's low mass IMF from the Quest-2 survey]{Mass function for all spectroscopically confirmed members presented
in this work. 
Thick-lined hatched histogram 
indicates all stars in our spectroscopic sample; the dotted
open histogram represents the same sample corrected for 
incomplete selection bins (\S\ref{cha:7:sec:usco:sub:imf}). 
}
\label{fig:cha7uscoimf}
\end{center}
\end{figure}

In Figure~\ref{fig:cha7uscoimf}, we show 
the IMF derived using the DM97 models for our spectroscopic survey of low mass stars
and brown dwarfs, covering the mass range 
$M\sim$0.2--0.02 M$_\odot$.  Because we have observed spectroscopically only $\sim$15\% of the photometric PMS star candidates,
before constructing a mass function, 
it is necessary to determine that the spectroscopic sample
presented here is representative of the USco low mass population as a whole.   We use a method similar to that employed
by \cite{2004ApJ...610.1045S} and determine the relative completeness of our sample  
by computing the number of stars observed spectroscopically compared to the number of photometric candidates present 
in uniformly-spaced bins perpendicular to the line of selection at the 1\% data contour (see Figure~\ref{fig:cha3cmdselect}).
As can be seen, 
the number of observed stars falls off considerably for stars with magnitudes 
$r\gtrsim$19.5 which could be observed only under the best seeing conditions.
The spectroscopic survey completeness peaks at $\sim$20\%.
We correct the IMF to uniform completeness at this level by determining the mass distribution of stars within each magnitude
bin.  We then add stars to the relevant mass bins based on the fractional completeness
of the magnitude bins they came from, relative to the maximum 20\% completeness level.

In order to extend our sample to higher masses, 
we combined masses derived for our sample with those for stars from the survey by \cite{2002AJ....124..404P}.
We chose to use the \cite{2002AJ....124..404P} sample because (1) it constitutes the largest sample of spectroscopically confirmed
low mass stellar members, and (2) candidates were selected from an $I,R-I$ CMD in a manner similar to
the methods employed to select our sample thus providing a complementary dataset. 
We specifically chose not to include the higher mass X-ray selected sources \citep{1999AJ....117.2381P} due to the 
large difference in selection techniques.
To provide the most analogous samples possible, we re-derived masses for the 160 stars in the \cite{2002AJ....124..404P} 
survey with M spectral types by first converting published spectral types and 2MASS magnitudes to temperature and luminosity
using the techniques described in \S\ref{cha:6:sec:discussion:sub:hr}, and then using the DM97 tracks and the same interpolation program used to generate masses for the sample of stars presented here.    
In combining samples from multiple sources, one must be careful to account for relative completeness.
\cite{2002AJ....124..404P} estimate spectroscopic completeness levels of 87\% for stars with 0.2 $< M/M_\odot <$0.8
and 67\% for stars with mass $<$0.2 M$_\odot$, whereas, our corrected survey is only 20\% complete
across all sampled mass bins.  
The number of stars in the 
\cite{2002AJ....124..404P} survey were thus multiplied by either (20\% complete/87\% complete) for stars with 0.2 $< M/M_\odot <$0.8,
or (20\% complete/67\% complete) for stars with $M<$0.2 M$_\odot$.
The two surveys also cover very different areas.  \cite{2002AJ....124..404P} looked at 9 deg$^2$ whereas
the Quest-2 survey looked at $\sim$150 deg$^2$.  However, as discussed in Paper~I, SCH06, and \cite{mythesis}, 
the Quest-2 survey coverage within
this area is not complete due to several failed CCDs, gaps between the CCDs, and incomplete $r-i$ color coverage. 
To account for differences in area, we scaled the IMF derived from the \cite{2002AJ....124..404P} data to match, on average, the level of the IMF 
derived from stars in our survey in the -0.6$>$log(M/M$_\odot$)$>$-1.0 mass bins.  
Due to the nature of magnitude limited surveys, the lowest
mass bin (or bins, depending on bin-width) of any IMF is usually incomplete relative to the rest of the derived
distribution; thus, for the IMF in the combined samples, we use the values derived from the Quest-2 data to 
populate the -1.0$>$log(M/M$_\odot$)$>$-1.4
bins of the IMF.

Figure~\ref{fig:cha7uscoimffull} shows IMFs for the low mass stellar and substellar population of the USco association. 
In total, the combined IMF contains $\sim$377 stars with masses as high as $\sim$0.6 M$_\odot$ and as low as $<$0.02 M$_\odot$.
The derived mass function rises with a slope of $dN/dM \propto M^{-1.13}$, peaks at $\sim$0.13 M$_\odot$, remains high
with a secondary peak at M$\sim$0.05 M$_\odot$, and then gradually falls off through the substellar regime.
Thus, we find the spectroscopic IMF for USco's low
mass population to be very different from the photometric IMF derived by 
\cite{2007MNRAS.374..372L} which had a flatter slope $dN/dM \propto M^{-0.3}$ 
from 0.3$ < M/M_\odot <$0.01, 
and peaked at $M\sim$0.01 M$_\odot$. 
The total mass inferred from the spectroscopic IMF over the mass range 0.6--0.02 M$_\odot$ is $\sim$48 M$_\odot$.  
If we correct this number to account for the
fact that the IMF shown here is only 20\% complete over $\sim$40\% of the entire spatial extent of the Hipparcos stars, 
the total mass inferred for association members less massive than $\sim$0.6 M$_\odot$ is $\sim$600 M$_\odot$.

\subsection{Comparisons of the Low Mass IMF between Star-Forming Regions}
\label{cha:7:sec:comp}

Diagnostic studies of stellar populations in
different locations and at varying stages of evolution 
are needed to explore the possibility of a universal mass function for low mass objects.
While one might expect that the IMF should vary with 
star formation environment, we do not yet have enough evidence
to determine if such a variation exists.  
Aside from the work presented here, 
numerous studies have been carried out 
to characterize the low mass stellar and substellar 
mass functions of other young clusters and associations in a variety of 
environments.  
Because of the intrinsic faintness of low mass objects, most surveys are photometric.
Authors then use a combination of theoretical models and statistical
analysis to transform a group's color-magnitude diagram or 
luminosity function into a photometric IMF which may not accurately represent the underlying population. 
Spectroscopic surveys provide a more accurate
assessment of the stellar membership and thus, of the underlying cluster population. 

The substellar populations
of several other young
star-forming regions have been studied spectroscopically
using techniques similar to those presented here, and we discuss results for USco in comparison to three other regions.  
\cite{2003ApJ...593.1093L} studied the rich cluster IC 348 in Perseus.
There have been several 
studies of the Taurus region, namely those by
\cite{2000ApJ...544.1044L}, \cite{2002ApJ...580..317B}, 
\cite{2003ApJ...590..348L}, and \cite{2004ApJ...617.1216L}, which together constitute a complete sample
within a selected region.
\cite{2004ApJ...610.1045S} used near-infrared imaging and spectroscopic observations to constrain the low mass IMF in the 
$\sim$1 Myr-old Orion Nebula Cluster (ONC).  
The data for IC 348 \citep{2003ApJ...593.1093L} and Taurus (\citealt{2000ApJ...544.1044L}, \citealt{2002ApJ...580..317B}, 
\citealt{2003ApJ...590..348L}, and \citealt{2004ApJ...617.1216L}) are published as complete across all mass ranges within the areas surveyed.  
\cite{2004ApJ...610.1045S} employed similar techniques to those used here to correct the derived IMF to a uniform 
40\% completeness across all magnitude ranges.
IC 348, Taurus, and the ONC are all young ($age\lesssim$3 Myr), and, similar to USco,
have not yet had time to lose members through dynamical ejection. 
Therefore, if the low mass IMF is universal, similar mass distributions
should be observed in all four regions.  

Our goal in this section is to assess the physical differences between stars formed in different star-forming environments.
As described in \S\ref{cha:6:sec:discussion:sub:hr}, 
effective temperatures shown in Figure~\ref{fig:cha6hrusco} are derived directly from observed
spectral types using a theoretical or (in our case) empirical temperature scale.
Observational evidence suggests that late-type young stars will have a similar mass across a large range of ages.
For example, similar masses of $M\sim$0.03--0.09 M$_\odot$ have been measured for M6--M7 binary or triple compenents 
in both the ONC ($\sim$1 Myr; \cite{2006Natur.440..311S})
and AB Dor ($\sim$75 Myr; \cite{2007ApJ...665..736C}).
Thus, spectral type distributions should provide a reasonable approximation to mass distributions for young low mass stars.
We therefore begin this exercise by first comparing spectral type distributions found in different star forming regions, and thereby avoid
introducing uncertainties that inherently arise when deriving stellar masses.

Spectral type distributions are shown in Figure~\ref{fig:cha7imfalldm}.
All four distributions are complete in a relative sense across spectral type bins for M-type stars.
Distributions for members of the ONC and members of USco discovered with Quest-2 were corrected to a uniform completeness level  
in a manner similar to that
described in \S\ref{cha:7:sec:usco:sub:imf}.   
We have included 
in the IC 348 and Taurus spectral-type distributions all M-type stars present in the IMF of the respective discovery paper.
As can be seen, the distribution for USco is visually dissimilar to the other regions, even considering the small sample sizes and large error bars.  This difference could be due to real variation in the IMF, but could also
be influenced by the difference in age of a few million years ($\sim$5 Myr vs. $\lesssim$3 Myr) between USco and the other regions.
We performed $\chi^2$tests between all four distributions to determine the probabilities that the distributions could have be drawn from the same population.
The tests yielded a 
very small probability ($\sim$10$^{-22}$) that the USco and ONC distributions could be drawn from the same population, and marginally-small 
probabilities at the 5--10\% levels of USco arising from the same population as either Taurus or IC 348.
However, we caution against taking results from the $\chi^2$ test at face value.  The histograms for USco and the ONC have been scaled to account
for incompleteness in magnitude and spatial area during the observations, and correctly and rigorously accounting for these scaling factors in the $\chi^2$ 
test is a non-trivial and not straight-forward exercise.   The Taurus and IC348 populations, however, have not been scaled, and have an almost 100\% chance of being drawn from the same population.  While this result argues
for a common spectral-type distribution between these two regions amoung the lowest mass stars and brown dwarfs, substantial variation was found by \cite{2003ApJ...593.1093L} for hotter, K-type stars.

We now compare directly mass distributions for the four selected star forming regions.
As discussed above and in \S\ref{cha:6:sec:discussion:sub:hr},
the mass of a star is inferred from placement on an HR diagram using a set 
of theoretical mass tracks, and the exact method by which the mass is derived can vary significantly from study to study.
Thus, direct comparison
of the published mass data itself requires great caution
 given that different mass tracks, temperature scales, and interpolation methods were used
for different studies.  
Quantitative assessment of the differences between theoretical PMS star models is beyond the scope of this work, as is
analysis of differences between the numerous temperature scales employed
by various authors who study young low mass stars.
We have used published spectral types and 2MASS $J,H,K_S$ magnitudes to derive effective temperatures, luminosities, and hence masses in 
a manner consistent with that used in our work in the ONC and USco.  
We reiterate that masses for all stars presented in this section were derived in a consistent manner;  however, variations in the
theoretical or empirical scales/tracks used could shift interpretation of the data.

IMFs derived from the DM97 tracks are shown in Figures~\ref{fig:cha7imfalldmmass}.
We find that the ONC, IC 348, and Taurus IMFs exhibit a fall-off of stars beyond $\sim$0.1 M$_\odot$.
The mass distribution for USco, however, does not begin to turnover until $\sim$0.05 M$_\odot$. 
Thus, similar to the spectral type distribution, the IMF derived for USco shows an over-abundance of
late-type, low mass stars, and suggests that the difference seen in USco's spectral type distribution did arise from a differnce in it's IMF, rather than from slight differences in age between USco and the other regions.  
Our work extends the claim by \cite{2002AJ....124..404P} that USco may contain anomolously 
high numbers (compared to the field) of low mass stars to lower masses, and is suggestive that USco
also contains relatively higher numbers of very low mass stars and brown dwarfs compared to other nearby star-forming regions. 

USco differs from IC 348, Taurus and the ONC in that it contains $\sim$50 very massive, O- and B-type stars, almost three times the number found in the ONC
(none are found in either Taurus or IC 348).  Thus, the conclusion that USco may contain relatively 
more very low mass stars than other young regions may be related
to its large number of high mass stars.
For example, 
\cite{2006ApJ...641..504A} have shown through numerical simulation that disks around lower mass stars are more
susceptible to destruction (further out
than $\sim$10 AU) from dynamical interactions with surrounding stars of higher mass.
It is possible that in a similar manner, low mass cores moving through a giant molecular cloud are more 
susceptible to stripping of their accretion envelope in the presence of higher mass cores.
Our results suggest that, within a star forming region, 
either the presence of large numbers of very massive stars, or the environmental conditions that lead to 
numerous massive
star formation,
may play a large role in determining the low mass IMF.
We note that this conclusion has been drawn from results for only four different environments.  Many more 
regions must be studied before a definitive explanation of the low mass IMF can be derived.

\section{Summary}
\label{summary}

We have completed a large-area $g,r,i$ photometric survey of $\sim$150 deg$^2$ in and near the
Upper Scorpius region of recent star formation.  From these data, combined with
2MASS near-infrared magnitudes, we selected candidate new PMS association members.  We present here spectral observations for
a total of 243 candidates observed at either Palomar or CTIO, from which we determined 145 ($\sim$60\%) 
to be bona fide new
Upper Scorpius members. 
We measured H$\alpha$ emission for all new members and determine 15 of the 145
low mass stars and brown dwarfs in the 5 Myr USco association 
are still accreting.  Based on comparison of the spatial distributions of low and high mass association
members, we find no evidence for spatial segregation in USco within the northern portion of the association. 

We used photometry and spectral types to derive effective temperatures
and luminosities, and placed newly identified association members onto an HR
diagram.  These data were combined with 
pre-main sequence evolutionary models to derive a mass and age for each star.  Using Monte Carlo simulations
we showed that, taking into account known observational errors, the observed age dispersion for the low
mass population in USco is consistent with all stars forming in a single burst $\sim$5 Myr ago,
and place an upper limit of $\pm$3 Myr (i.e., 60\%) on the age spread if
the star formation rate has been constant in time.
We also derived the first spectroscopic mass function for USco that extends into the substellar regime
and compared these results to those for three other young clusters and associations.

\section{Appendix A}
Magnitudes for the Quest-2 photometric candidates spectroscopically determined to be field stars are listed in Table~3.

$\;$

{\it Facilities:} \facility{PO:1.2m (Quest-2)}, \facility{Hale (Double Spectrograph)}, \facility{Blanco (Hydra)}

\nocite{2004AJ....127..449M}
\nocite{2002AJ....124..404P}
\nocite{1997ApJS..112..109B}
\nocite{2000AJ....120..479A}
\nocite{1987AJ.....94..106W}

\bibliography{../../apjmnemonic,../../thesis}

\begin{thebibliography}{47}
\expandafter\ifx\csname natexlab\endcsname\relax\def\natexlab#1{#1}\fi

\bibitem[{{Adams} {et~al.}(2006){Adams}, {Proszkow}, {Fatuzzo}, \&
  {Myers}}]{2006ApJ...641..504A}
{Adams}, F.~C., {Proszkow}, E.~M., {Fatuzzo}, M., \& {Myers}, P.~C. 2006, ApJ,
  641, 504

\bibitem[{{Ardila} {et~al.}(2000){Ardila}, {Mart{\'{\i}}n}, \&
  {Basri}}]{2000AJ....120..479A}
{Ardila}, D., {Mart{\'{\i}}n}, E., \& {Basri}, G. 2000, AJ, 120, 479

\bibitem[{{Baltay} {et~al.}(2007){Baltay}, {Rabinowitz}, {Andrews}, {Bauer},
  {Ellman}, {Emmet}, {Hudson}, {Hurteau}, {Jerke}, {Lauer}, {Silge},
  {Szymkowiak}, {Adams}, {Gebhard}, {Musser}, {Doyle}, {Petrie}, {Smith},
  {Thicksten}, \& {Geary}}]{2007PASP..119.1278B}
{Baltay}, C., {Rabinowitz}, D., {Andrews}, P., {Bauer}, A., {Ellman}, N.,
  {Emmet}, W., {Hudson}, R., {Hurteau}, T., {Jerke}, J., {Lauer}, R., {Silge},
  J., {Szymkowiak}, A., {Adams}, B., {Gebhard}, M., {Musser}, J., {Doyle}, M.,
  {Petrie}, H., {Smith}, R., {Thicksten}, R., \& {Geary}, J. 2007, \pasp, 119,
  1278

\bibitem[{{Baraffe} {et~al.}(1998){Baraffe}, {Chabrier}, {Allard}, \&
  {Hauschildt}}]{1998A&A...337..403B}
{Baraffe}, I., {Chabrier}, G., {Allard}, F., \& {Hauschildt}, P.~H. 1998, A\&A,
  337, 403

\bibitem[{{Barrado y Navascu{\'e}s} \&
  {Mart{\'{\i}}n}(2003)}]{2003AJ....126.2997B}
{Barrado y Navascu{\'e}s}, D. \& {Mart{\'{\i}}n}, E.~L. 2003, AJ, 126, 2997

\bibitem[{{Barsony} {et~al.}(1997){Barsony}, {Kenyon}, {Lada}, \&
  {Teuben}}]{1997ApJS..112..109B}
{Barsony}, M., {Kenyon}, S.~J., {Lada}, E.~A., \& {Teuben}, P.~J. 1997, ApJS,
  112, 109

\bibitem[{{Bessell} \& {Brett}(1988)}]{1988PASP..100.1134B}
{Bessell}, M.~S. \& {Brett}, J.~M. 1988, PASP, 100, 1134

\bibitem[{{Brice{\~n}o} {et~al.}(2002){Brice{\~n}o}, {Luhman}, {Hartmann},
  {Stauffer}, \& {Kirkpatrick}}]{2002ApJ...580..317B}
{Brice{\~n}o}, C., {Luhman}, K.~L., {Hartmann}, L., {Stauffer}, J.~R., \&
  {Kirkpatrick}, J.~D. 2002, ApJ, 580, 317

\bibitem[{{Carpenter}(2000)}]{2000AJ....120.3139C}
{Carpenter}, J.~M. 2000, AJ, 120, 3139

\bibitem[{{Close} {et~al.}(2007){Close}, {Thatte}, {Nielsen}, {Abuter},
  {Clarke}, \& {Tecza}}]{2007ApJ...665..736C}
{Close}, L.~M., {Thatte}, N., {Nielsen}, E.~L., {Abuter}, R., {Clarke}, F., \&
  {Tecza}, M. 2007, \apj, 665, 736

\bibitem[{{D'Antona} \& {Mazzitelli}(1997)}]{1997MmSAI..68..807D}
{D'Antona}, F. \& {Mazzitelli}, I. 1997, Memorie della Societa Astronomica
  Italiana, 68, 807

\bibitem[{{de Bruijne}(1999)}]{1999MNRAS.310..585D}
{de Bruijne}, J.~H.~J. 1999, MNRAS, 310, 585

\bibitem[{{de Zeeuw} {et~al.}(1999){de Zeeuw}, {Hoogerwerf}, {de Bruijne},
  {Brown}, \& {Blaauw}}]{1999AJ....117..354D}
{de Zeeuw}, P.~T., {Hoogerwerf}, R., {de Bruijne}, J.~H.~J., {Brown}, A.~G.~A.,
  \& {Blaauw}, A. 1999, AJ, 117, 354

\bibitem[{{Gomez} {et~al.}(1993){Gomez}, {Hartmann}, {Kenyon}, \&
  {Hewett}}]{1993AJ....105.1927G}
{Gomez}, M., {Hartmann}, L., {Kenyon}, S.~J., \& {Hewett}, R. 1993, AJ, 105,
  1927

\bibitem[{{Guieu} {et~al.}(2006){Guieu}, {Dougados}, {Monin}, {Magnier}, \&
  {Mart{\'{\i}}n}}]{2006A&A...446..485G}
{Guieu}, S., {Dougados}, C., {Monin}, J.-L., {Magnier}, E., \& {Mart{\'{\i}}n},
  E.~L. 2006, A\&A, 446, 485

\bibitem[{{Haisch} {et~al.}(2001){Haisch}, {Lada}, \&
  {Lada}}]{2001ApJ...553L.153H}
{Haisch}, Jr., K.~E., {Lada}, E.~A., \& {Lada}, C.~J. 2001, ApJ, 553, L153

\bibitem[{{Hillenbrand}(2005)}]{2005astro.ph.11083H}
{Hillenbrand}, L.~A. 2005, astro-ph, 0511083

\bibitem[{{Hillenbrand} {et~al.}(2008){Hillenbrand}, {Bauermeister}, \&
  {White}}]{2008ASPC..384..200H}
{Hillenbrand}, L.~A., {Bauermeister}, A., \& {White}, R.~J. 2008, in
  Astronomical Society of the Pacific Conference Series, Vol. 384, 14th
  Cambridge Workshop on Cool Stars, Stellar Systems, and the Sun, ed.
  S.~P.~P.~S.~D.~E.~M.~B.~J. {Messina}, 200

\bibitem[{{Hillenbrand} \& {White}(2004)}]{2004ApJ...604..741H}
{Hillenbrand}, L.~A. \& {White}, R.~J. 2004, ApJ, 604, 741

\bibitem[{{Kraus} {et~al.}(2008){Kraus}, {Ireland}, {Martinache}, \&
  {Lloyd}}]{2008ApJ...679..762K}
{Kraus}, A.~L., {Ireland}, M.~J., {Martinache}, F., \& {Lloyd}, J.~P. 2008,
  \apj, 679, 762

\bibitem[{{Kraus} {et~al.}(2005){Kraus}, {White}, \&
  {Hillenbrand}}]{2005ApJ...633..452K}
{Kraus}, A.~L., {White}, R.~J., \& {Hillenbrand}, L.~A. 2005, ApJ, 633, 452

\bibitem[{{Lada} \& {Lada}(2003)}]{2003ARA&A..41...57L}
{Lada}, C.~J. \& {Lada}, E.~A. 2003, ARA\&A, 41, 57

\bibitem[{{Lada} \& {Lada}(1995)}]{1995AJ....109.1682L}
{Lada}, E.~A. \& {Lada}, C.~J. 1995, AJ, 109, 1682

\bibitem[{{Leggett} {et~al.}(1996){Leggett}, {Allard}, {Berriman}, {Dahn}, \&
  {Hauschildt}}]{1996ApJS..104..117L}
{Leggett}, S.~K., {Allard}, F., {Berriman}, G., {Dahn}, C.~C., \& {Hauschildt},
  P.~H. 1996, ApJS, 104, 117

\bibitem[{{Leggett} {et~al.}(2002){Leggett}, {Golimowski}, {Fan}, {Geballe},
  {Knapp}, {Brinkmann}, {Csabai}, {Gunn}, {Hawley}, {Henry}, {Hindsley},
  {Ivezi{\'c}}, {Lupton}, {Pier}, {Schneider}, {Smith}, {Strauss}, {Uomoto}, \&
  {York}}]{2002ApJ...564..452L}
{Leggett}, S.~K., {Golimowski}, D.~A., {Fan}, X., {Geballe}, T.~R., {Knapp},
  G.~R., {Brinkmann}, J., {Csabai}, I., {Gunn}, J.~E., {Hawley}, S.~L.,
  {Henry}, T.~J., {Hindsley}, R., {Ivezi{\'c}}, {\v Z}., {Lupton}, R.~H.,
  {Pier}, J.~R., {Schneider}, D.~P., {Smith}, J.~A., {Strauss}, M.~A.,
  {Uomoto}, A., \& {York}, D.~G. 2002, ApJ, 564, 452

\bibitem[{{Lodieu} {et~al.}(2008){Lodieu}, {Hambly}, {Jameson}, \&
  {Hodgkin}}]{2008MNRAS.383.1385L}
{Lodieu}, N., {Hambly}, N.~C., {Jameson}, R.~F., \& {Hodgkin}, S.~T. 2008,
  \mnras, 383, 1385

\bibitem[{{Lodieu} {et~al.}(2007){Lodieu}, {Hambly}, {Jameson}, {Hodgkin},
  {Carraro}, \& {Kendall}}]{2007MNRAS.374..372L}
{Lodieu}, N., {Hambly}, N.~C., {Jameson}, R.~F., {Hodgkin}, S.~T., {Carraro},
  G., \& {Kendall}, T.~R. 2007, MNRAS, 374, 372

\bibitem[{{Loinard} {et~al.}(2008){Loinard}, {Torres}, {Mioduszewski}, \&
  {Rodr{\'{\i}}guez}}]{2008ApJ...675L..29L}
{Loinard}, L., {Torres}, R.~M., {Mioduszewski}, A.~J., \& {Rodr{\'{\i}}guez},
  L.~F. 2008, ApJ, 675, L29

\bibitem[{{Luhman}(2000)}]{2000ApJ...544.1044L}
{Luhman}, K.~L. 2000, ApJ, 544, 1044

\bibitem[{{Luhman}(2004)}]{2004ApJ...617.1216L}
---. 2004, ApJ, 617, 1216

\bibitem[{{Luhman} {et~al.}(2003{\natexlab{a}}){Luhman}, {Brice{\~n}o},
  {Stauffer}, {Hartmann}, {Barrado y Navascu{\'e}s}, \&
  {Caldwell}}]{2003ApJ...590..348L}
{Luhman}, K.~L., {Brice{\~n}o}, C., {Stauffer}, J.~R., {Hartmann}, L., {Barrado
  y Navascu{\'e}s}, D., \& {Caldwell}, N. 2003{\natexlab{a}}, ApJ, 590, 348

\bibitem[{{Luhman} {et~al.}(2003{\natexlab{b}}){Luhman}, {Stauffer}, {Muench},
  {Rieke}, {Lada}, {Bouvier}, \& {Lada}}]{2003ApJ...593.1093L}
{Luhman}, K.~L., {Stauffer}, J.~R., {Muench}, A.~A., {Rieke}, G.~H., {Lada},
  E.~A., {Bouvier}, J., \& {Lada}, C.~J. 2003{\natexlab{b}}, ApJ, 593, 1093

\bibitem[{{Mamajek} {et~al.}(2004){Mamajek}, {Meyer}, {Hinz}, {Hoffmann},
  {Cohen}, \& {Hora}}]{2004ApJ...612..496M}
{Mamajek}, E.~E., {Meyer}, M.~R., {Hinz}, P.~M., {Hoffmann}, W.~F., {Cohen},
  M., \& {Hora}, J.~L. 2004, ApJ, 612, 496

\bibitem[{{Mart{\'{\i}}n} {et~al.}(2004){Mart{\'{\i}}n}, {Delfosse}, \&
  {Guieu}}]{2004AJ....127..449M}
{Mart{\'{\i}}n}, E.~L., {Delfosse}, X., \& {Guieu}, S. 2004, AJ, 127, 449

\bibitem[{{Massey} {et~al.}(1988){Massey}, {Strobel}, {Barnes}, \&
  {Anderson}}]{1988ApJ...328..315M}
{Massey}, P., {Strobel}, K., {Barnes}, J.~V., \& {Anderson}, E. 1988, ApJ, 328,
  315

\bibitem[{{Preibisch} {et~al.}(2002){Preibisch}, {Brown}, {Bridges},
  {Guenther}, \& {Zinnecker}}]{2002AJ....124..404P}
{Preibisch}, T., {Brown}, A.~G.~A., {Bridges}, T., {Guenther}, E., \&
  {Zinnecker}, H. 2002, AJ, 124, 404

\bibitem[{{Preibisch} {et~al.}(2001){Preibisch}, {Guenther}, \&
  {Zinnecker}}]{2001AJ....121.1040P}
{Preibisch}, T., {Guenther}, E., \& {Zinnecker}, H. 2001, AJ, 121, 1040

\bibitem[{{Preibisch} {et~al.}(1998){Preibisch}, {Guenther}, {Zinnecker},
  {Sterzik}, {Frink}, \& {Roeser}}]{1998A&A...333..619P}
{Preibisch}, T., {Guenther}, E., {Zinnecker}, H., {Sterzik}, M., {Frink}, S.,
  \& {Roeser}, S. 1998, A\&A, 333, 619

\bibitem[{{Preibisch} \& {Zinnecker}(1999)}]{1999AJ....117.2381P}
{Preibisch}, T. \& {Zinnecker}, H. 1999, AJ, 117, 2381

\bibitem[{{Preibisch} \& {Zinnecker}(2007)}]{2007IAUS..237..270P}
{Preibisch}, T. \& {Zinnecker}, H. 2007, in IAU Symposium, Vol. 237, IAU
  Symposium, ed. B.~G. {Elmegreen} \& J.~{Palous}, 270--277

\bibitem[{{Rabinowitz} {et~al.}(2003){Rabinowitz}, {Baltay}, {Emmet},
  {Hurteau}, {Snyder}, {Andrews}, {Ellman}, {Morgan}, {Bauer}, {Musser},
  {Gebhard}, {Adams}, {Djorgovski}, {Mahabal}, {Graham}, {Bogosavljevic},
  {Williams}, {Brucato}, \& {Thicksten}}]{2003AAS...203.3812R}
{Rabinowitz}, D., {Baltay}, C., {Emmet}, W., {Hurteau}, T., {Snyder}, J.,
  {Andrews}, P., {Ellman}, N., {Morgan}, N., {Bauer}, A., {Musser}, J.,
  {Gebhard}, M., {Adams}, G., {Djorgovski}, G., {Mahabal}, A., {Graham}, M.,
  {Bogosavljevic}, M., {Williams}, R., {Brucato}, R., \& {Thicksten}, R. 2003,
  in Bulletin of the American Astronomical Society, 1262

\bibitem[{{Salpeter}(1955)}]{1955ApJ...121..161S}
{Salpeter}, E.~E. 1955, ApJ, 121, 161

\bibitem[{{Shu} {et~al.}(1987){Shu}, {Adams}, \&
  {Lizano}}]{1987ARA&A..25...23S}
{Shu}, F.~H., {Adams}, F.~C., \& {Lizano}, S. 1987, ARA\&A, 25, 23

\bibitem[{{Skrutskie} {et~al.}(2006){Skrutskie}, {Cutri}, {Stiening},
  {Weinberg}, {Schneider}, {Carpenter}, {Beichman}, {Capps}, {Chester},
  {Elias}, {Huchra}, {Liebert}, {Lonsdale}, {Monet}, {Price}, {Seitzer},
  {Jarrett}, {Kirkpatrick}, {Gizis}, {Howard}, {Evans}, {Fowler}, {Fullmer},
  {Hurt}, {Light}, {Kopan}, {Marsh}, {McCallon}, {Tam}, {Van Dyk}, \&
  {Wheelock}}]{2006AJ....131.1163S}
{Skrutskie}, M.~F., {Cutri}, R.~M., {Stiening}, R., {Weinberg}, M.~D.,
  {Schneider}, S., {Carpenter}, J.~M., {Beichman}, C., {Capps}, R., {Chester},
  T., {Elias}, J., {Huchra}, J., {Liebert}, J., {Lonsdale}, C., {Monet}, D.~G.,
  {Price}, S., {Seitzer}, P., {Jarrett}, T., {Kirkpatrick}, J.~D., {Gizis},
  J.~E., {Howard}, E., {Evans}, T., {Fowler}, J., {Fullmer}, L., {Hurt}, R.,
  {Light}, R., {Kopan}, E.~L., {Marsh}, K.~A., {McCallon}, H.~L., {Tam}, R.,
  {Van Dyk}, S., \& {Wheelock}, S. 2006, AJ, 131, 1163

\bibitem[{{Slesnick}(2007)}]{mythesis}
{Slesnick}, C.~L. 2007, PhD thesis, California Institute of Technology

\bibitem[{{Slesnick} {et~al.}(2006a){Slesnick}, {Carpenter}, \&
  {Hillenbrand}}]{2006AJ....131.3016S}
{Slesnick}, C.~L., {Carpenter}, J.~M., \& {Hillenbrand}, L.~A. 2006a, AJ, 131, 3016 (Paper~I) 

\bibitem[{{Slesnick} {et~al.}(2006b){Slesnick}, {Carpenter},
  {Hillenbrand}, \& {Mamajek}}]{2006AJ...132.2665S}
{Slesnick}, C.~L., {Carpenter}, J.~M., {Hillenbrand}, L.~A., \& {Mamajek}, E.E. 2006b, AJ, 132, 2665 (SCH06) 

\bibitem[{{Slesnick} {et~al.}(2004){Slesnick}, {Hillenbrand}, \&
  {Carpenter}}]{2004ApJ...610.1045S}
{Slesnick}, C.~L., {Hillenbrand}, L.~A., \& {Carpenter}, J.~M. 2004, ApJ, 610,
  1045

\bibitem[{{Stassun} {et~al.}(2006){Stassun}, {Mathieu}, \&
  {Valenti}}]{2006Natur.440..311S}
{Stassun}, K.~G., {Mathieu}, R.~D., \& {Valenti}, J.~A. 2006, \nat, 440, 311

\bibitem[{{Wilking} {et~al.}(1987){Wilking}, {Schwartz}, \&
  {Blackwell}}]{1987AJ.....94..106W}
{Wilking}, B.~A., {Schwartz}, R.~D., \& {Blackwell}, J.~H. 1987, AJ, 94, 106

\end{thebibliography}
\bibliographystyle{apj}

\clearpage 

\begin{figure}
\begin{center}
\scalebox{0.55}{\includegraphics{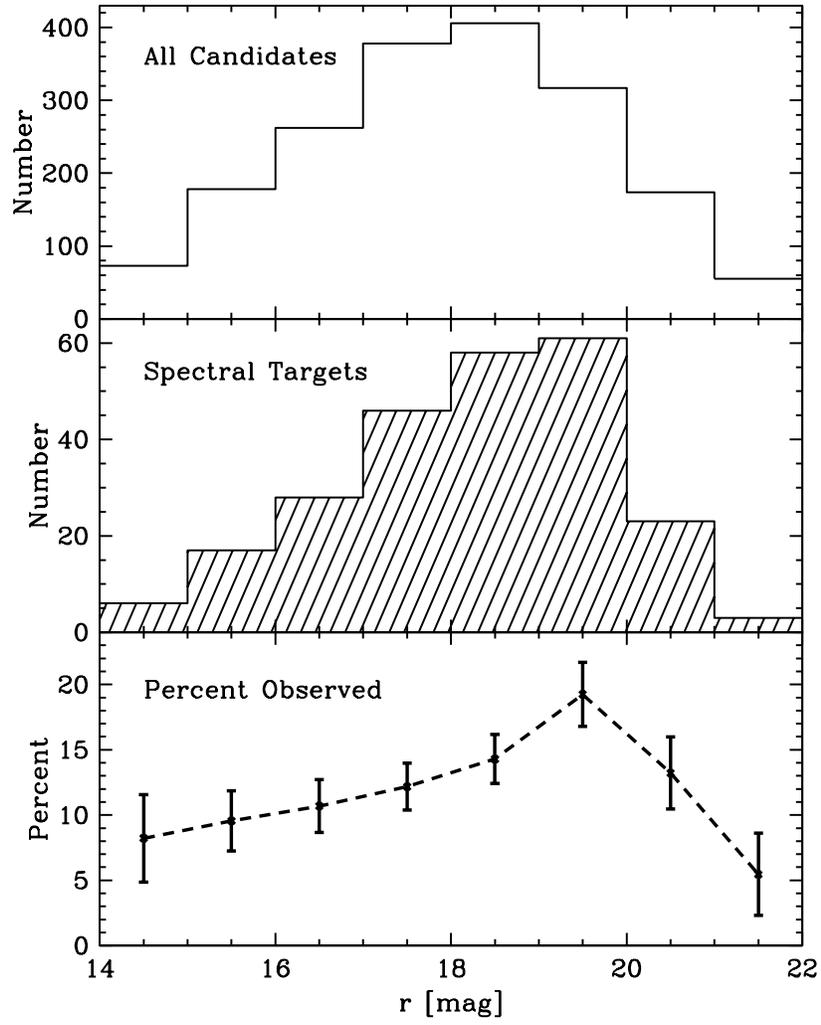}}
\caption[Histogram of USco magnitudes]{Histogram of optical CMD completeness as a function of $r$ magnitude.
Bin centers correspond to $r$ magnitudes at the 1\% contour line in Figure~\ref{fig:cha3cmdselect}.  
The top panel shows data for all photometric PMS candidates selected as in \S\ref{obs:phot},
and the middle panel shows
data for all targets observed spectroscopically.  
The bottom panel shows the percent completeness with $\sqrt{N}$ errorbars.  
Completeness peaks at 20\% near $r$=19.5 mag.
}
\label{fig:cha7maghistusco}
\end{center}
\end{figure}
 
\begin{figure}
\begin{center}
\rotatebox{-90}{
\scalebox{0.5}{\includegraphics{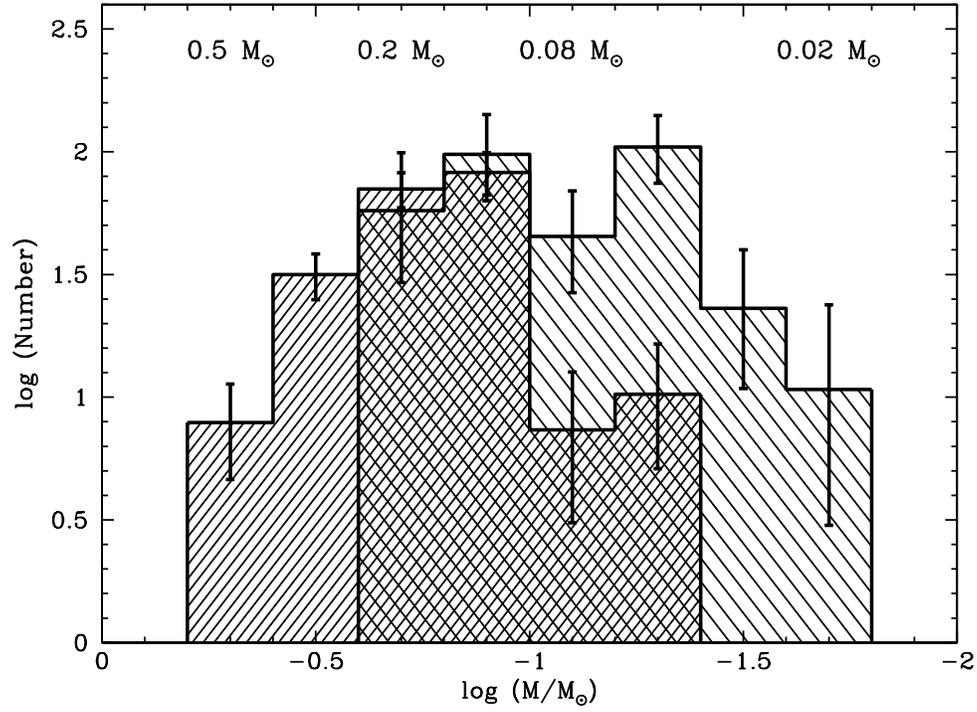}}}
\caption[USco's full low mass IMF]{IMF for the USco association from $\sim$0.6--0.02 M$_\odot$.  
The light and dark hatched histograms depict the IMFs derived from the Quest-2 survey and the 
\cite{2002AJ....124..404P} survey, respectively.  Both IMFs have been scaled to a uniform 20\% completeness,
and data from the \cite{2002AJ....124..404P} survey have been scaled to match
the data presented here in the -0.6$>$log(M/M$_\odot$)$>$-1.0 mass bins.  All histograms
are shown with $\sqrt{N}$ errorbars.
}
\label{fig:cha7uscoimffull}
\end{center}
\end{figure}

\begin{figure}
\begin{center}
\scalebox{0.6}{\includegraphics{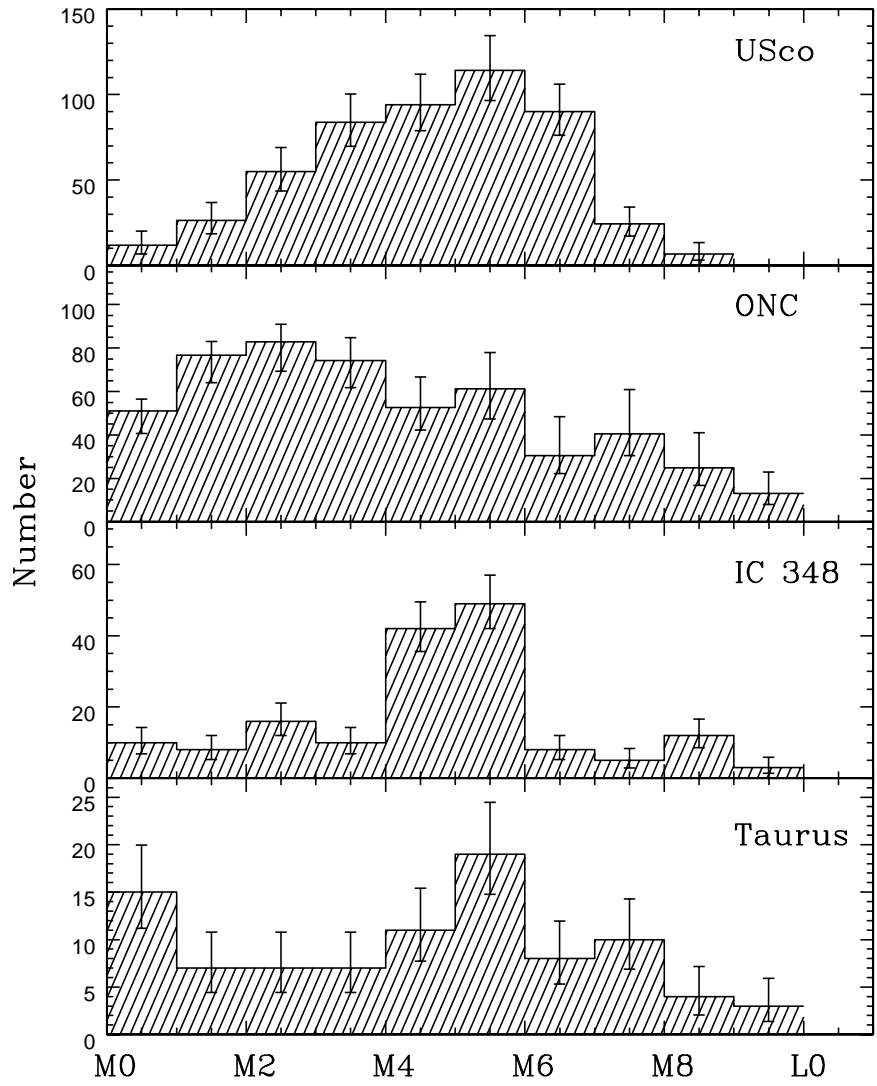}}
\caption[Comparison of spectral type distributions between four different star forming regions]
{Low-mass spectral type distributions for USco, the ONC, IC 348, and Taurus.  Data for the latter
three regions were taken from the literature (see text).  
}
\label{fig:cha7imfalldm}
\end{center}
\end{figure}
 
\begin{figure}
\begin{center}
\scalebox{0.6}{\includegraphics{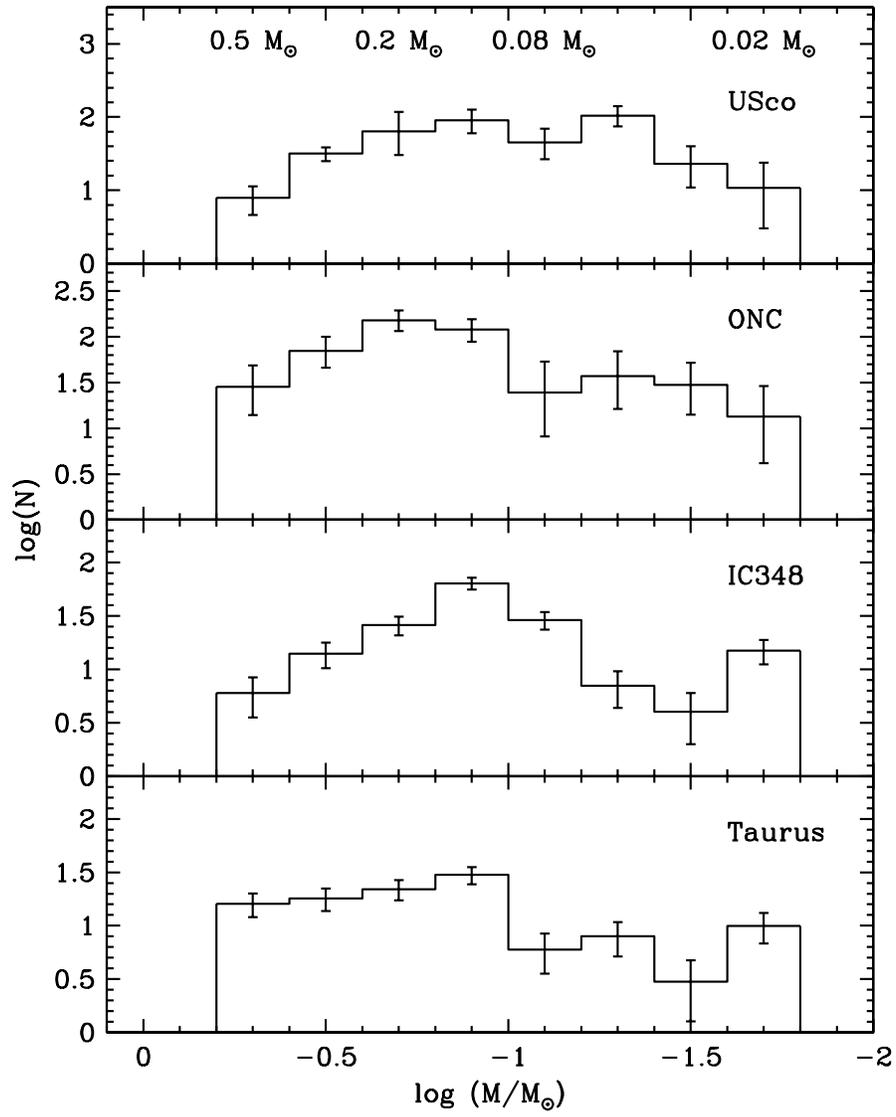}}
\caption[Comparison of IMFs for star forming regions using DM97 tracks]
{Low mass IMFs for USco, the ONC, IC 348, and Taurus generated from DM97 mass tracks.  Data for the latter
three regions were taken from the literature (see text).  
}
\label{fig:cha7imfalldmmass}
\end{center}
\end{figure}

\end{document}